\documentclass{vgtc} 

\usepackage{mathptmx}
\usepackage{graphicx}
\usepackage{times}
\usepackage{amsmath}

\usepackage[normalsize]{subfigure}

\usepackage{cite} 
\usepackage[final]{hyperref} 
\hypersetup{
	colorlinks=true,       
	linkcolor=blue,        
	citecolor=blue,        
	filecolor=magenta,     
	urlcolor=blue         
}

\onlineid{0}

\vgtccategory{Research}

\title{Higher-order Hermite-Gauss modes as a robust flat beam in interferometric gravitational wave detectors}

\author{Liu Tao$^{1}$\thanks{Corresponding author: liu.tao@ligo.org} %
\and Anna Green$^{1}$ %
\and Paul Fulda$^{1}$}
\affiliation{\scriptsize $^{1}$University of Florida, 2001 Museum Road, Gainesville, Florida 32611, USA}

\abstract{Higher-order Laguerre-Gauss (LG) modes have previously been investigated as a candidate for reducing test-mass thermal noise in ground-based gravitational-wave detectors like Advanced LIGO. It has been shown however that LG modes' fragility against mirror surface figure imperfections limits their compatibility with the current state-of-the-art test masses. In this paper we explore the alternative of using higher-order Hermite-Gauss (HG) modes for thermal noise reduction, and show that with the deliberate addition of astigmatism they are orders of magnitude more robust against mirror surface distortions than LG modes of equivalent order. We present simulations of Advanced LIGO-like arm cavities with realistic mirror figures which can support HG$_{33}$ modes with average arm losses and contrast defects in a Fabry-Perot Michelson interferometer configuration which are well below the typical measured values in Advanced LIGO. This demonstrates that the mirror surface flatness errors will not be a limiting factor for the use of these modes in future gravitational-wave detectors. } 

\begin{document}

\maketitle
\section{Introduction}

The sensitivity of 
all leading gravitational wave detectors 
is limited at signal frequencies around 100\,Hz by the thermal noise of the test masses~\cite{aLIGO,AdVirgo}. A major goal of the gravitational-wave community is therefore to reduce the effect of this noise. It has been proposed to use laser beams with a more uniform intensity distribution than the fundamental Gaussian beam in order to better `average out' the effects of this thermal noise \cite{ Mours_2006, Vinet_2007}. In particular, research into the potential of the Laguerre-Gauss (LG) mode LG$_{33}$ has been carried out using numerical simulations and tabletop experiments~\cite{PhysRevD.79.122002, PhysRevD.82.012002}. 
However, it has been shown that the surface distortions present even in state-of-the-art mirrors will cause significant impurity and losses for the LG$_{33}$ mode in realistic, high finesse cavities \cite{PhysRevD.84.102002, PhysRevD.84.102001, Sorazu}. While at first glance the thermal noise benefit afforded by higher-order Hermite-Gauss (HG) modes is more modest than that of LG modes, they have other properties that may make them more suitable for use in laser interferometers.
It is the aim of this paper to investigate the robustness of higher-order HG modes and LG modes of the same order against mirror surface deformations by numerical simulations performed using \textsc{Finesse}~\cite{Freise_2004, finesse}. In particular, we investigate the performance of the HG$_{33}$ and LG$_{22}$ modes in aLIGO-like linear cavities and in a Fabry-Perot Michelson interferometer. We also discuss the possibility of using odd-indexed HG modes with segmented mirrors, given their property of having intensity nulls along the principal axes. 

The paper is structured as follows: we give a short introduction about higher-order Hermite-Gauss modes and their thermal noise benefits in Section~\ref{sec:HGmodes}. Section~\ref{sec:model} then describes the interferometer model that is used to perform the simulations reported in this paper. In Section~\ref{sec:initialresults} we report the results of simulations for the HG$_{33}$, LG$_{22}$ and HG$_{00}$ modes in terms of the relevant figures of merit: arm cavity loss, arm mode purity and contrast defect.
In Section~\ref{sec:rotation}, we demonstrate that HG$_{33}$ performance can be slightly improved by rotating the mirrors such as to minimize oblique astigmatism. In Section~\ref{sec:astigmatism}, we show how deliberately \emph{increasing} the vertical astigmatism in mirrors causes a dramatic improvement in the HG$_{33}$ performance in terms of mode loss, purity and contrast defect reduction. We report our conclusions and discuss prospects for further study in Section~\ref{sec:conclusion}. Appendix~\ref{sec:randommaps} includes a description of the process of creating the random realistic test mass surface figures which are used in the simulations.

\section{Hermite-Gauss modes}
\label{sec:HGmodes}
For any paraxial beam propagating along the z axis, the spatial profile in the transverse orthogonal x and y directions can be expanded in the Hermite-Gauss mode basis. The HG modes 
are a complete and orthonormal set of functions defined by mode indices n and m. The spatial order of the mode is defined as the sum n + m. The general expression for the spatial distribution of the mode HG$_\mathrm{nm}$ can be given as~\cite{Bond2017}:
\begin{equation}
\label{equ:HG}
U_{\mathrm{nm}}(x, y, z)=U_{\mathrm{n}}(x, z) U_{\mathrm{m}}(y, z),
\end{equation}
with
\begin{equation}
\begin{aligned}
U_{\mathrm{n}}(x, z)=&\left(\frac{2}{\pi}\right)^{1 / 4}\left(\frac{\exp (\mathrm{i}(2 \mathrm{n}+1) \Psi(z))}{2^\mathrm{n} \mathrm{n} ! w(z)}\right)^{1 / 2} \\
& \times H_\mathrm{n}\left(\frac{\sqrt{2} x}{w(z)}\right) \exp \left(-\mathrm{i} \frac{k x^{2}}{2 R_{C}(z)}-\frac{x^{2}}{w^{2}(z)}\right),
\end{aligned}
\label{equ:HGnm}
\end{equation}
where $H_\mathrm{n}(x)$ is the Hermite polynomial of order n, $k$ is the wavenumber, and $w(z)$, $Rc(z)$ and $\Psi(z)$ are the beam radius, wavefront radius of curvature and Gouy phase respectively, as commonly defined for the fundamental Gaussian beam.

Higher-order Hermite-Gauss modes are good candidates for thermal noise reduction. The thermal noise performance for higher-order HG modes was calculated for fixed beam size $w$ by Vinet~\cite{PhysRevD.82.042003}. Here we focus on the coating Brownian noise, because this is the dominant test mass thermal noise source in gravitational wave detectors. The coating Brownian noise power spectral density improvement factors $\Theta_\mathrm{n m}^{\mathrm{CB}}$ for HG$_\mathrm{nm}$ modes over the HG$_{00}$ mode are shown in Tab.~\ref{tab:thermalnoise}, where for each mode the beam size is scaled to maintain 1\,ppm clipping loss on a fixed-radius circular mirror.

\begin{table}[htbp]
\centering
\caption{ Improvement in coating Brownian noise power spectral density, $\Theta_\mathrm{n m}^{\mathrm{CB}}$ for HG$_\mathrm{nm}$ modes compared to the HG$_{00}$ mode.  All modes are scaled to give 1\,ppm clipping loss on a fixed sized circular mirror.}
\begin{tabular}{c||cccccc}
n $\backslash$ m & 0 & 1 & 2 & 3 & 4 & 5 \\
\hline \hline 0 & 1 & 1.10 & 1.11 & 1.08 & 1.05 & 1.02 \\
1 & 1.10 & 1.29 & 1.33 & 1.40 & 1.30 & 1.27 \\
2 & 1.10 & 1.33 & 1.40 & 1.41 & 1.41 & 1.39 \\
3 & 1.08 & 1.32 & 1.41 & 1.44 & 1.45 & 1.45 \\
4 & 1.05 & 1.30 & 1.41 & 1.45 & 1.47 & 1.47 \\
5 & 1.02 & 1.27 & 1.39 & 1.45 & 1.47 & 1.48 \\
\hline \hline
\end{tabular}
\label{tab:thermalnoise}
\end{table}

The observable volume of space, and therefore the rate of detection of homogeneously distributed gravitational wave sources, is roughly proportional to the inverse cube of the detector noise amplitude spectral density. 
In particular, the detection rate improvement factor for HG$_{33}$ mode over HG$_{00}$ mode, $R_{33}$ is therefore
\begin{equation}
    \text R_{33} = (\Theta_{3 3}^{\mathrm{CB}})^{3/2}=1.44^{3/2} = 1.73,
    \label{equ:decectionRates}
\end{equation}
if we assume the detector is only limited by thermal noise. In this paper we will focus on the HG$_{33}$ mode as a higher-order HG mode example. HG modes of higher order do have some additional thermal noise benefit, but the benefits diminish quickly beyond the HG$_{33}$ mode, as shown in Tab.~\ref{tab:thermalnoise}. The conclusions drawn about the HG$_{33}$ mode performance are not limited to the HG$_{33}$ mode in particular, but can reasonably be extrapolated to other HG modes as well.

Future interferometers are likely to use silicon test masses, have longer arms, and require larger and more massive mirrors~\cite{voyager,cosmicx}. 
Currently we are up against the industry technology limitation for the diameter of circular `boules' of high purity silicon~\cite{voyager}. 
We can however imagine fabricating larger mirrors from multiple substrates, and utilizing high-order Hermite-Gauss modes for the readout beam because they can be arranged to have intensity nulls at the bonding lines (where the thermal noise is likely to be high and optical quality may be low).

\begin{figure}
    \centering
    \includegraphics[width=0.95\linewidth]{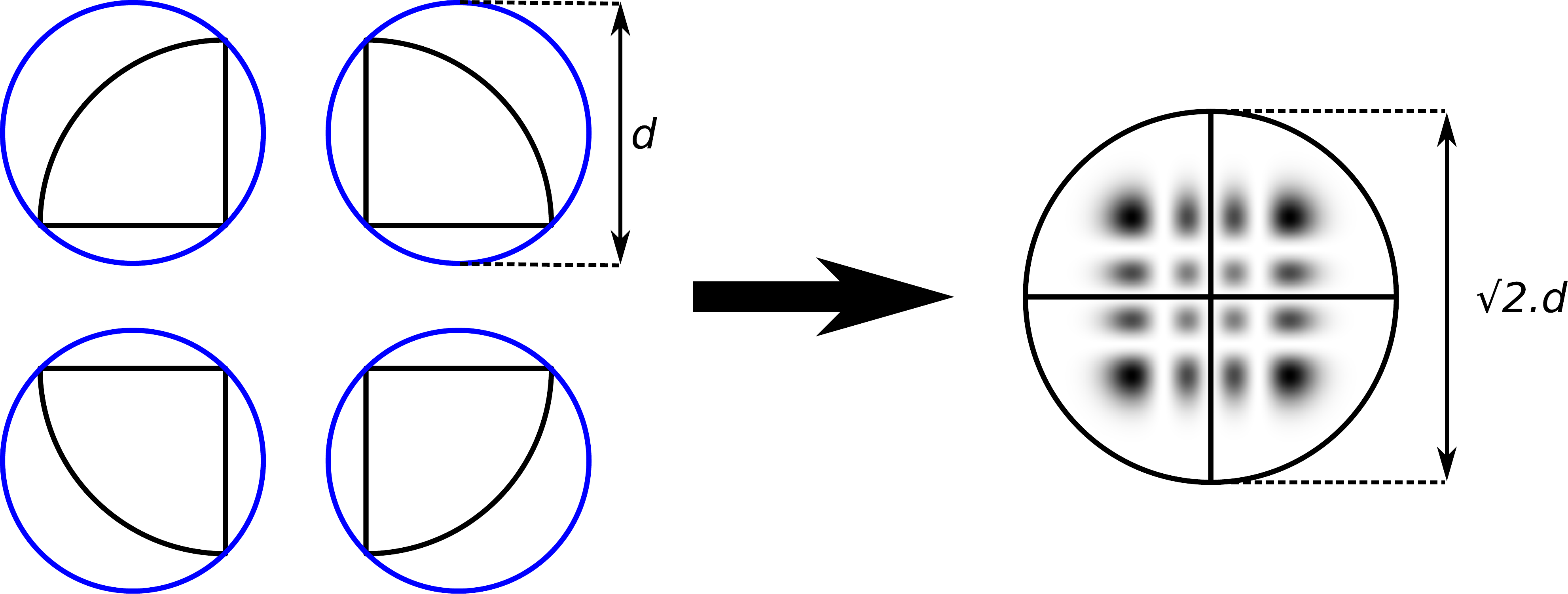}
    \caption{Illustration of a HG$_{33}$ mode incident on a circular segmented mirror. Assuming the same clipping loss, the beam size increases by a factor of $\sqrt{2}$.}
    \label{fig:compound_mirror_circular}
\end{figure}

Fig.~\ref{fig:compound_mirror_circular} shows an example implementation. The segmented mirror is formed by combining four quadrants. The bonding lines are arranged such that they are lined up with the intensity nulls of the HG$_{33}$ mode. The radius of the compound mirror is $\sqrt{2}$ larger than the original mirror, which allows further reduction of thermal noises by supporting larger beam sizes.
The coating thermal noise power spectral density, for example, is inversely proportional to the square of the beam size at the mirror. 

The larger segmented mirror therefore provides an additional improvement factor of 2 in addition to the factor 1.44 shown in Tab.~\ref{tab:thermalnoise} for the HG$_{33}$ mode, which leads to an increased detection rate:
\begin{equation}
    \text R_{33}^\mathrm{Seg} = (1.44\times2)^{3/2} = 4.89.
\end{equation}
Quantum radiation pressure noise is expected to be another limiting noise source at lower frequencies in future detectors. As a force noise its impact on the strain sensitivity of detectors scales inversely with the mass of the mirrors, which itself scales with the square of the mirror diameter (or even the cube if we maintain fixed relative dimensions). The compatibility of the HG$_{33}$ mode with segmented mirrors can also therefore lead to a reduction in this limiting noise source.

\section{The Model}
\label{sec:model}
\label{subsec:opticallayout}
\begin{figure}[htbp]
    \centering
    \includegraphics[width=\linewidth]{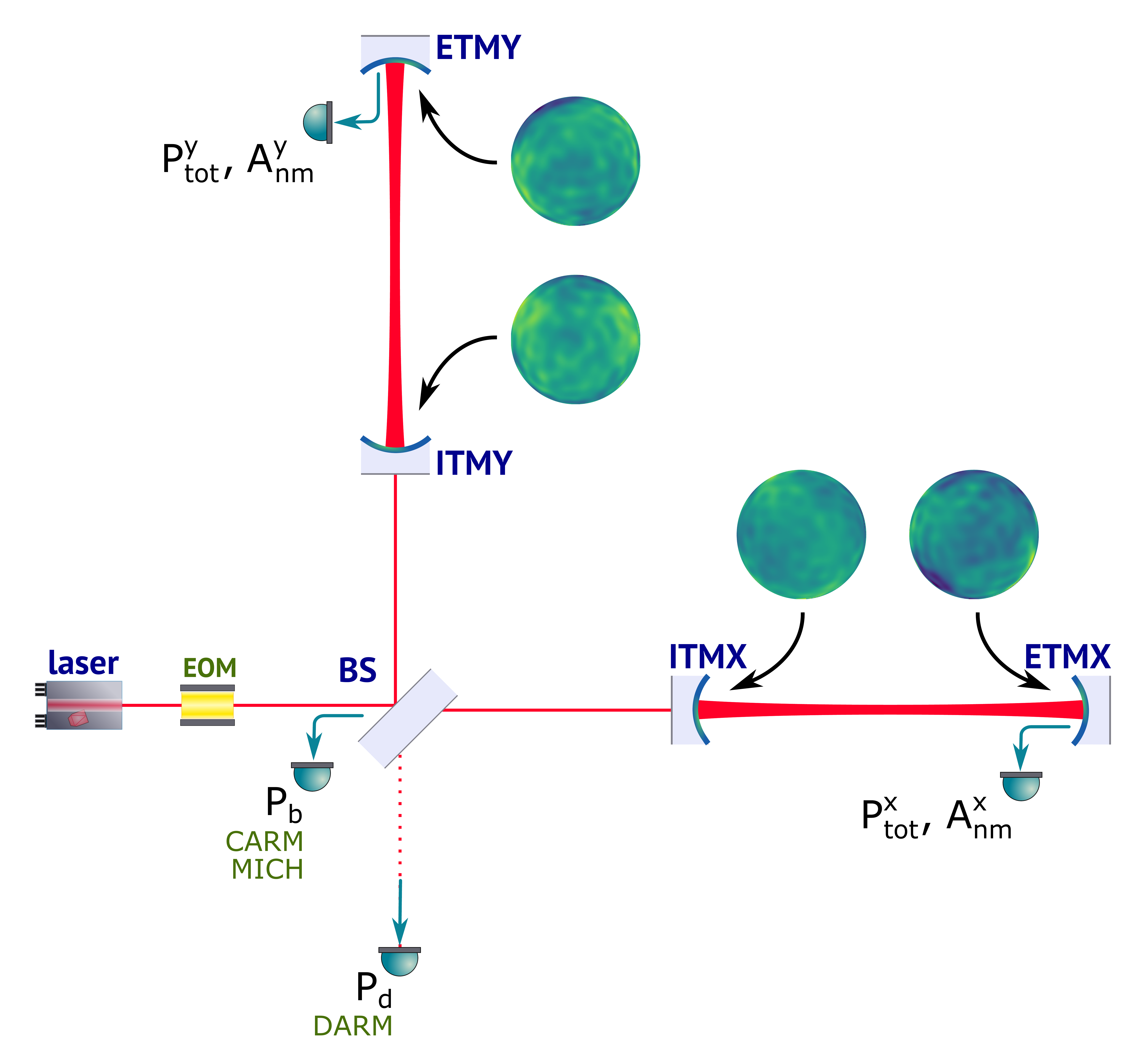}
    \caption{The optical model. The interferometer under test is a Fabry-Perot Michelson interferometer, with different realistic mirror surface figures applied to each of the 4 cavity mirrors.}

    \label{fig:interferometer}
\end{figure}

The optical model for our study is an aLIGO-like Fabry-Perot Michelson interferometer, as shown in Fig.~\ref{fig:interferometer}. Four different `maps', describing mirror surface figure imperfections, are applied to each of the test masses. Using this model we look at three figures of merit: the individual loss and mode impurity in each arm, and the contrast defect of the Fabry-Perot Michelson~\footnote{The optical layout~\ref{fig:interferometer} involves two cavities already so we will get two data points for the loss and impurity for each contrast defect data point.}. Although the true aLIGO configuration also includes a power recycling mirror and resonant sideband extraction mirror, these are not expected to significantly impact the results for the aforementioned figures of merit.

The higher-order HG and LG mode performances are calculated numerically using the interferometer simulation tool \textsc{Finesse}~\cite{Freise_2004, finesse} with its python wrapper \textsc{Pykat}~\cite{pykat,brown2020pykat}, which is a commonly used simulation software in the gravitational-wave instrument science community. For this study we compare the HG$_{33}$ mode with the LG$_{22}$ mode. This choice was made because both modes have the same order (LG$_\mathrm{pl}$ mode order is given as 2p+l), and equal indices. Furthermore, the HG$_{33}$ mode has intensity nulls on the principal axes, making it suitable for use with segmented mirrors. We also compare against the currently used HG$_{00}$ mode for reference, and typical aLIGO measured values where available

The parameters in the optical layout
are similar to the aLIGO design \cite{aLIGO}, with the exception of the test mass radii of curvature. These are instead symmetric, for simplicity, 
and are carefully chosen such that the clipping loss is always 1\,ppm for all mode cases. The radii of curvature for each mode used are listed in Tab.~\ref{tab:cavityparameters}.

\begin{table}[htbp]
\centering
\caption{ Radii of curvature for the arm cavity mirrors when using different input spatial modes.}
\begin{tabular}{lccc}
&HG$_{00}$ & HG$_{33}$ &LG$_{22}$\\\hline \hline 
ITM \(\mathrm{R}_{c} \) [m]  & -2091.67 & -2679.93 & -2789.58 \\
ETM \(\mathrm{R}_{c} \) [m] & -2091.67 & 2679.93 & 2789.58 \\
\hline \hline
\end{tabular}
\label{tab:cavityparameters}
\end{table}

Our goal is to assess the performance of the HG and LG modes in an interferometer with realistic mirror surface figures. There are however only a limited number of aLIGO measured test mass surface figures, or `maps', available. We therefore created our own randomized mirror maps in order to generate the large number ($\sim$4000) of aLIGO-representative maps that were needed to build statistics when analyzing the performance of the different laser modes. Each randomized mirror map is constructed in such a way as to have a spatial frequency spectrum which is similar to a measured aLIGO `base' map. For more detailed discussion of the randomized map generation process, see Appendix~\ref{sec:randommaps}.

Applying mirror surface figures to the test masses in the simulation can lead to a detuning of the length degrees of freedom from the optimal case (resonant cavities and dark Michelson fringe). In a realistic interferometer however, these detunings would be eliminated by the length control feedback loops. We therefore implemented a realistic RF modulation/demodulation scheme in our model using an electro-optic modulator (EOM) as shown in Fig.~\ref{fig:interferometer} at frequency of 60\,MHz to sense the three pertinent length degrees of freedom: common arm length (CARM), differential arm length (DARM) and Michelson tuning (MICH). CARM is defined as the summation of the two arm cavity lengths and is used to keep the arms on resonance. DARM is defined as the difference of the two arm cavity lengths and is used to get the best interference between the two arms at the dark port. MICH is defined as the difference of the two short arm lengths (between the ITMs and beam splitter) and it keeps the output port on a dark fringe. Mirror tunings were then adjusted in each simulation trial in order to keep the two arms resonant, and the Michelson on a dark fringe. We also use a very small modulation depth (0.0001) so sideband leakage to the dark port has insignificant contribution to contrast defect.

\section{Spatial mode performance}
\label{sec:initialresults}
\subsection{Mode loss and impurity}
\label{sec:initiallossandimpurity}

The mode loss with a pure HG$_\mathrm{nm}$ mode or LG$_\mathrm{nm}$ mode as the cavity input beam is defined as
\begin{equation}
\Lambda =1-\left(\frac{\mathrm{E}_\mathrm{nm}^{map}}{\mathrm{E}_\mathrm{nm}^{no\,map}}\right)^2
\end{equation}
where $\mathrm{E}_\mathrm{nm}^{map}$ and $\mathrm{E}_\mathrm{nm}^{no\,map}$ are the field amplitude of the HG$_\mathrm{nm}$ or LG$_\mathrm{nm}$ modes with and without maps applied respectively, as shown in Fig.~\ref{fig:interferometer}. 

The mode impurity with pure mode as input is defined as
\begin{equation}
 \Pi =1 - \frac{\left|\mathrm{E}_\mathrm{nm}\right|^{2}}{\text{P}_{\text {tot }}}
\end{equation} 
where $\text P_\mathrm{tot}$ is the total power inside the cavity, as shown in Fig.~\ref{fig:interferometer} as well.

In order to calculate the purity for LG$_{22}$ mode, it was necessary to calculate the LG$_{22}$ content. However, \textsc{Finesse} internally calculates transverse field profiles in the HG basis only, and thus post-processing was necessary to convert back to the LG basis to calculate the LG$_{22}$ content. This was achieved by combining the amplitudes of all HG modes of order 6 (HG$_{60}$, HG$_{51}$, HG$_{42}$,...,HG$_{06}$) with the appropriate coefficients given by the expression~\cite{Bond2017}
\begin{equation}
u_{p, l}^{L G}(x, y, z)=\sum_{k=0}^{N}(-1)^{p}(\mp \mathrm{i})^{k} b(|l|+p, p, k) u_{N-k, k}^{H G}(x, y, z)
\end{equation}
where $N=2 p+|l|$, $\pm$ is negative for positive l and positive for negative l and with real coefficients
\begin{equation}
b(n, m, k)=\sqrt{\frac{(N-k) ! k !}{2^{N} n ! m !}} \frac{1}{k !}\left(\partial_{t}\right)^{k}\left[(1-t)^{n}(1+t)^{m}\right]_{t=0}.
\end{equation}

The mode loss and mode impurity for HG$_{00}$, HG$_{33}$ and LG$_{22}$ modes with aforementioned aLIGO-like random maps applied were calculated in 1968 trials. Fig.~\ref{fig:rotationloss} and  Fig.~\ref{fig:rotationpurity} show the results for the loss and purity respectively. The averages and standard deviations are listed in Tab.~\ref{tab:avgstdlosspurity}. It shows that the HG$_{33}$ mode has marginally smaller average impurity and mode loss than the LG$_{22}$ mode from the their average values. Comparing to the fundamental HG$_{00}$ case, however, we see that the loss and impurity for HG$_{33}$ and LG$_{22}$ are several orders of magnitude larger.

\begin{table}[htbp]
\centering
\caption{Averages and standard deviations of loss and impurity for HG$_{00}$, HG$_{33}$ and LG$_{22}$ (1968 trials)}
\begin{tabular}{rrccc}
&&HG$_{00}$ & HG$_{33}$ & LG$_{22}$   \\
\hline\hline
Loss &Avg [ppm] &68.7&10973.9&18221.9\\
     & Std [ppm]&18.8& 11829.9& 13862.6\\ \hline
Impurity & Avg [ppm] &1.1&	5484.5&	9058.4\\
         & Std [ppm]&0.5&5982.6& 7010.9\\
\hline\hline
\end{tabular}
  \label{tab:avgstdlosspurity}
\end{table}

\subsection{Contrast defect}
\label{subsec:initialCD}

We calculated the contrast defect that results from applying four random maps to the four mirrors, for the three spatial modes under test. The contrast defect, $C$ was calculated as
\begin{equation}
    \text C = \frac{P_{d}}{P_{b}},
\end{equation}
where $P_{d}$ and $P_{b}$ are the total power measured at the dark port and bright port respectively, as shown in Fig.~\ref{fig:interferometer}.

Contrast defects were calculated in this way for HG$_{00}$, HG$_{33}$ and LG$_{22}$ modes for 984 trials. The result for the average and standard deviation are listed in Tab.~\ref{tab:avgstdcd}; full results for HG$_{33}$ and LG$_{22}$ are also plotted in Fig.~\ref{fig:cd10nm}. Similar to the loss and impurity cases (see Tab.~\ref{tab:avgstdlosspurity}), on average the contrast defect for the HG$_{33}$ mode is several orders of magnitudes larger than the currently used HG$_{00}$ mode in aLIGO, though it does have slightly smaller values compared to LG$_{22}$ mode.

\begin{table}[htbp]
\centering
\caption{Averages and standard deviations of the contrast defect for HG$_{00}$, HG$_{33}$ and LG$_{22}$ (984 trials)}
\begin{tabular}{rrccc}
&&HG$_{00}$ & HG$_{33}$ & LG$_{22}$   \\
\hline\hline
Contrast Defect & Avg [ppm] &1.7&	11795.8& 20026.7\\
                & Std [ppm]&0.9& 13584.5& 16666.7\\
\hline\hline
\end{tabular}
  \label{tab:avgstdcd}
\end{table}

\section{Rotating maps to reduce HG mode loss}
\label{sec:rotation}
\subsection{Individual Zernike contributions}

To gain a deeper understanding of the specific mirror surface features that contribute most to the mode loss and mode impurity of HG$_{33}$ and LG$_{22}$ modes, we simulated a cavity with mirror surfaces described by individual Zernike polynomial terms of 1\,nm amplitude.

\begin{figure}[htbp]
\centering
\includegraphics[width=\linewidth]{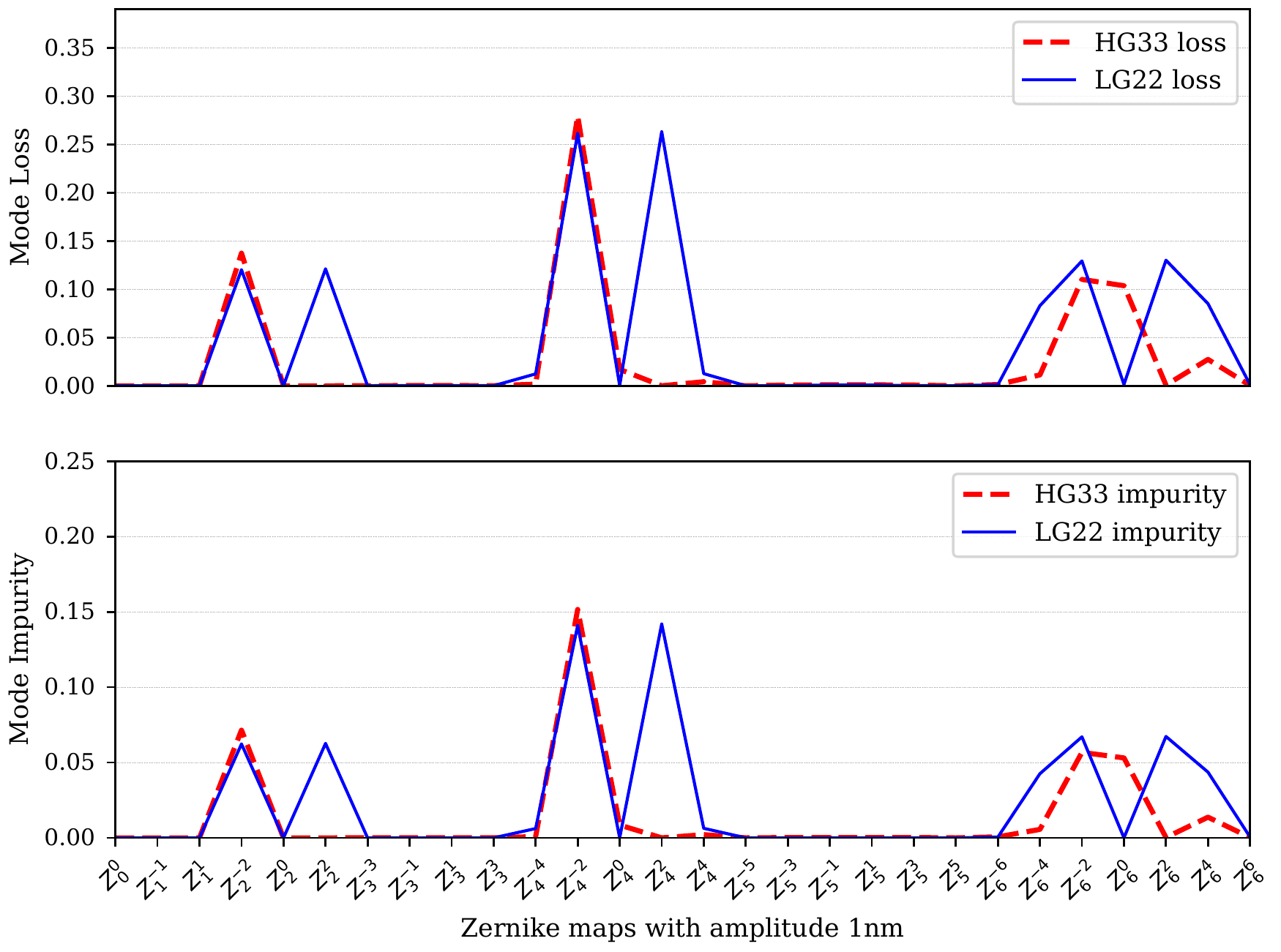}
\caption{The mode loss and impurity of the HG$_{33}$ and LG$_{22}$ modes for each individual Zernike term with amplitude 1\,nm.}
\label{fig:modelossandpurityLGtwHGt}
\end{figure}

Fig.~\ref{fig:modelossandpurityLGtwHGt} shows the mode loss and impurity, per Zernike term with 1\,nm amplitude, for the HG$_{33}$ and LG$_{22}$ modes. We see that the plots for mode loss and impurity are qualitatively almost identical, indicating that scattering of the HG$_{33}$ and LG$_{22}$ modes into pseudo-degenerate modes of the same order is the primary source of loss in these cases.
Fig.~\ref{fig:modelossandpurityLGtwHGt} also shows that the HG$_{33}$ mode is relatively impervious to the $Z^2_2$ (vertical astigmatism) and $Z^4_2$ (vertical secondary astigmatism) terms when compared with the LG$_{22}$ mode. 
This suggests that the mirror may be rotated to minimize the more problematic oblique astigmatism terms $Z_{2}^{-2}$ and $Z_{4}^{-2}$.

\subsection{The effect of rotating maps}
The $Z_{2}^{-2}$ term is identical to a $Z_{2}^{2}$ term rotated by 45$^{\circ}$. 
It therefore follows that we can rotate maps about their center such that the coefficient $A_{2}^{-2}$ in the new map is minimized, thus reducing the loss for the HG$_{33}$ mode.

The functional forms of the vertical and oblique astigmatism terms ($Z_{2}^{2}$ and $Z_{2}^{-2}$ respectively) are:
\begin{eqnarray}
    Z_{2}^{2} = \sqrt{6} \rho^{2} \cos 2 \theta \\
    Z_{2}^{-2} = \sqrt{6} \rho^{2} \sin 2 \theta,
\end{eqnarray}
where $\rho$ is the radial coordinate and $\theta$ is the azimuthal angle.

An arbitrary weighted combination of these two can then be rewritten as:
\begin{equation}
    S = A_{2}^{2} \rho^{2} \cos 2 \theta + A_{2}^{-2} \rho^{2} \sin 2 \theta.
\end{equation}
Since the Zernike polynomials are defined over a unit disk, 
we can equivalently write 
\begin{equation}\label{equ: rotationangle}
    S =  A_{rot} \rho^{2} \cos{2 (\theta + \alpha)},
\end{equation}
where $A_{rot}= \sqrt{(A_{2}^{2})^2 + (A_{2}^{-2})^2 }$ is the $Z_{2}^{2}$ coefficient after the rotation, and $\alpha = \arctan(A_{2}^{-2}/A_{2}^{2})/2$ is the rotation angle needed to minimize $A_{2}^{-2}$.

\begin{figure}[tbp]
\subfigure[Mode Loss (1968 trials)]{
    \label{fig:rotationloss}
    \includegraphics[width=\linewidth]{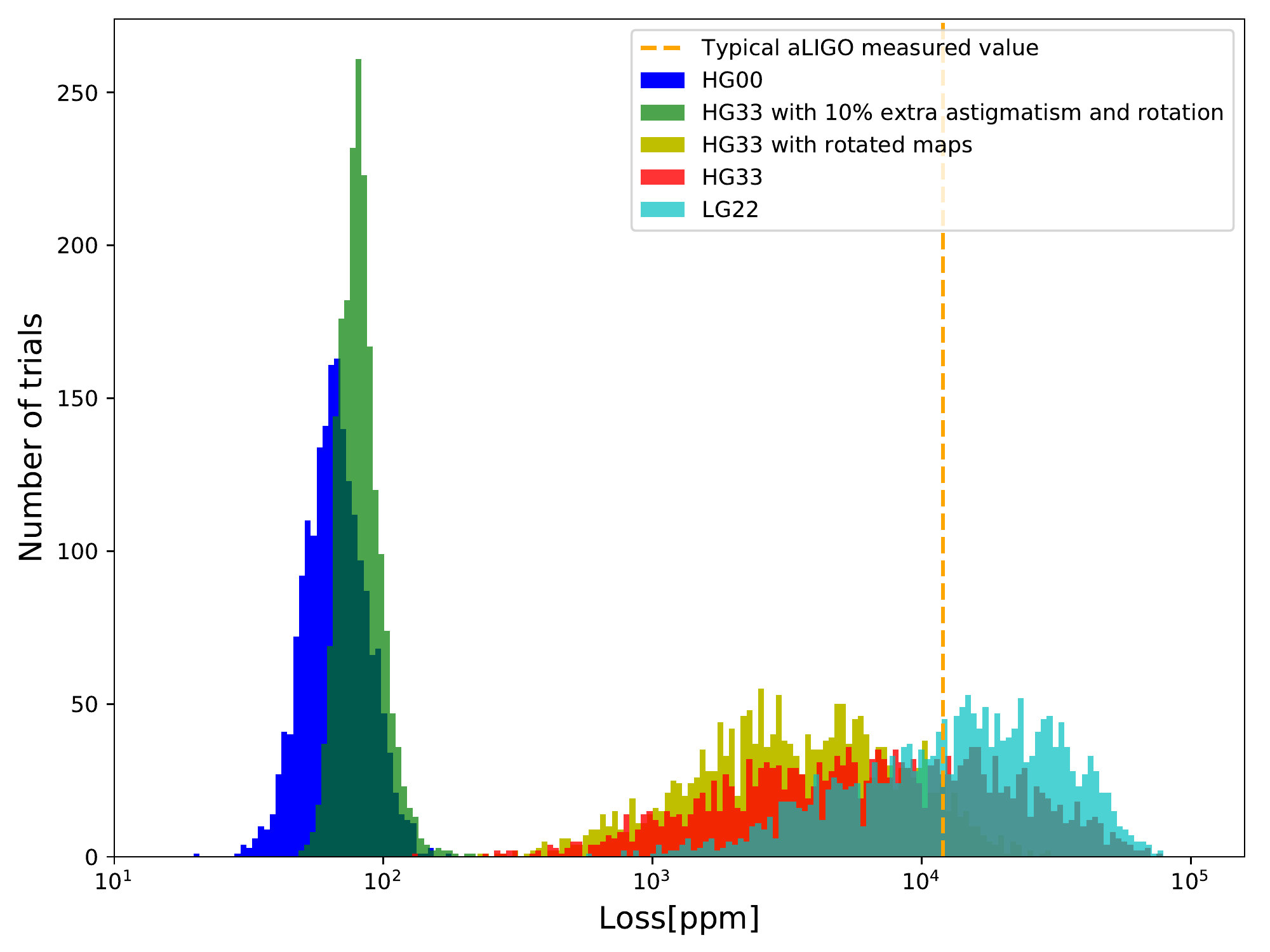}
}

\subfigure[Mode Impurity (1968 trials)]{
    \label{fig:rotationpurity}
    \includegraphics[width=\linewidth]{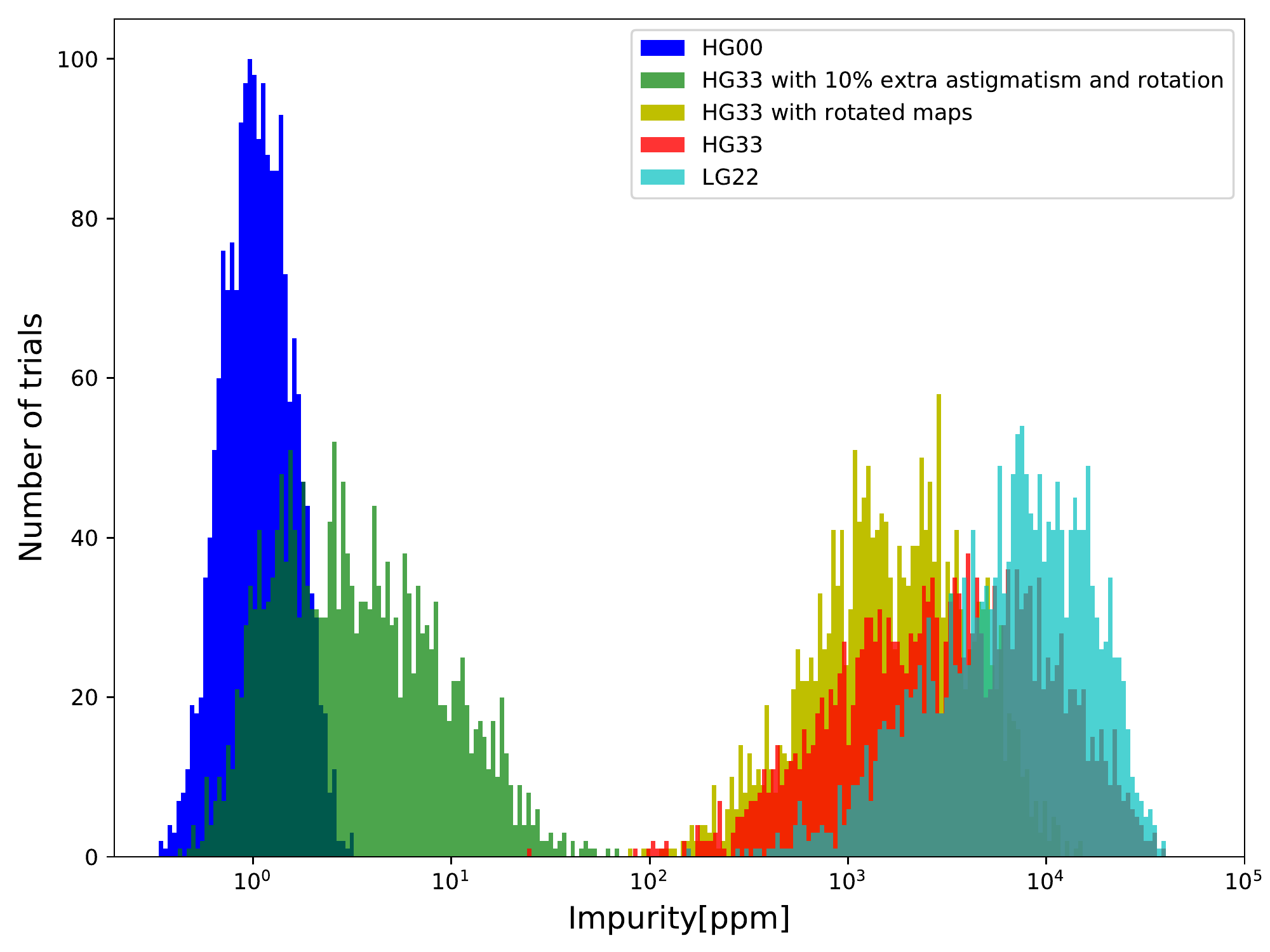}
}
\subfigure[Contrast Defect (984 trials)]{
    \label{fig:cd10nm}
    \includegraphics[width=\linewidth]{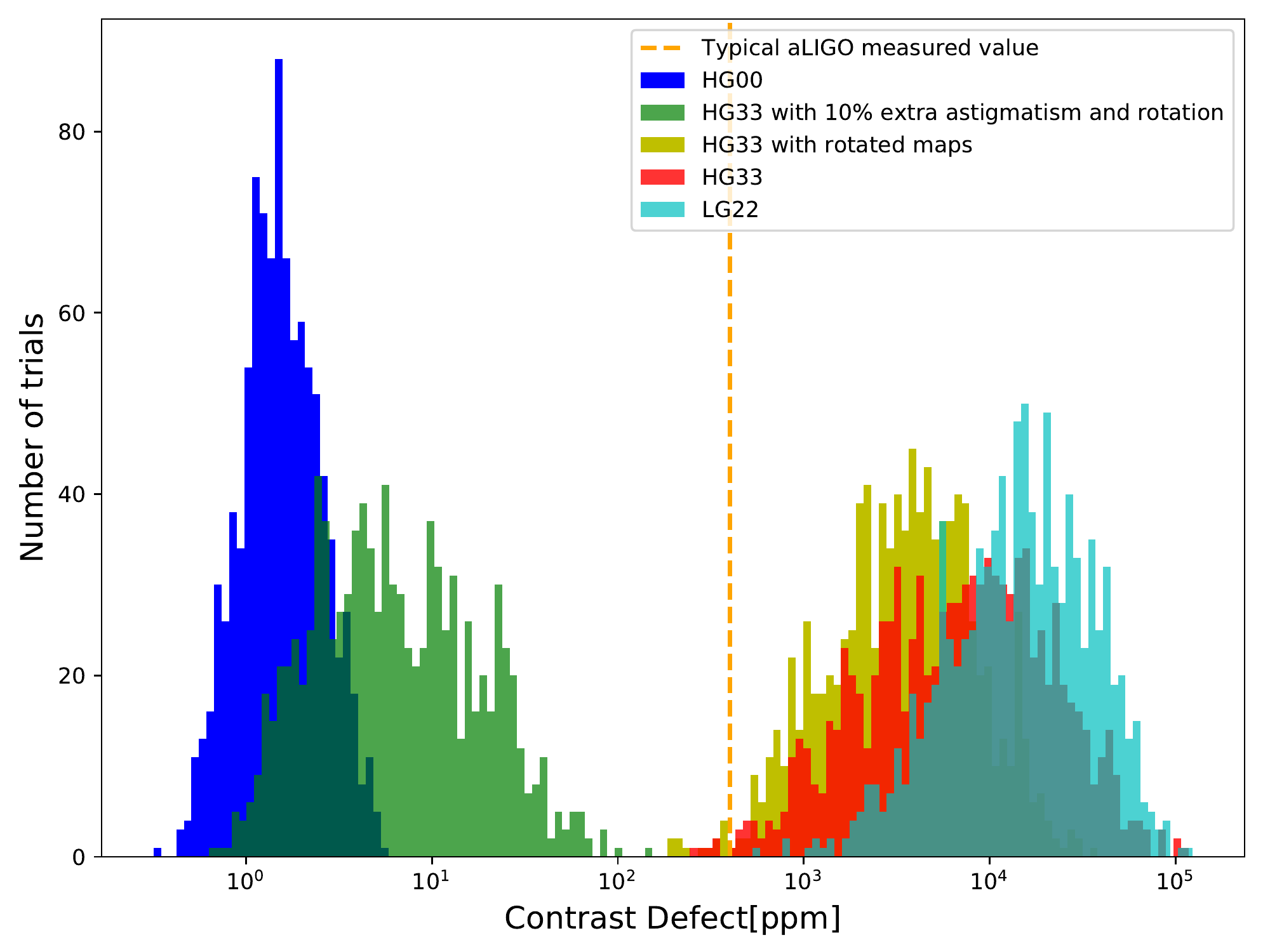}
}
\caption{Histograms showing the mode loss, impurity and contrast defect for 984 full interferometer simulations. Three HG$_{33}$ cases are shown: with the original maps, with maps rotated to minimize oblique astigmatism, and with 10\,\% astigmatism and subsequent rotation. The HG$_{00}$ case is shown for reference, along with typical aLIGO measured values where available.}
\label{fig:histograms}
\vspace{-0.16725pt}
\end{figure}

The effects of map rotation on the performance of the HG$_{33}$ modes in the simulated interferometer are shown in yellow in Fig.~\ref{fig:histograms}. 
For each random map, the rotation angle was calculated using Eqn.~\ref{equ: rotationangle}. As can be seen in the figure, the loss, impurity, and contrast defect are marginally reduced when compared to the original HG$_{33}$ case (red). The average for HG$_{33}$ loss, impurity, contrast defect after rotating the map is around 2 times smaller than the case before rotation (see Tab.~\ref{tab:avgstdcompensated}).

\begin{table}[htbp]
\centering
\caption{Averages and standard deviations of the mode loss, impurity, and contrast defect for the HG$_{33}$ mode when using rotated maps, and when applying 10\% extra astigmatism and then rotating.
(1968 trials; 984 for contrast defect)
}
\begin{tabular}{rrp{1.4cm}p{2cm}}
&& HG$_{33}$ \newline + Rotation & HG$_{33}$ \newline + Astigmatism \newline + Rotation\\
\hline\hline
Loss             & Avg [ppm] & 4803.7 & 82.9\\
                 & Std [ppm] & 4132.8 & 16.4\\ \hline
Impurity         & Avg [ppm] & 2371.1 & 5.3\\
                 & Std [ppm] & 2074.9 & 6.3\\\hline
Contrast Defect  & Avg [ppm] & 4873.1 & 10.5\\
                 & Std [ppm] & 4501.9 & 12.9\\
\hline\hline
\end{tabular}
  \label{tab:avgstdcompensated}
\end{table}

\section{Improvement by adding astigmatism}
\label{sec:astigmatism}

Eqn.~\ref{equ:HG} shows that HG modes are separable in the x and y axes. As such, HG modes can describe the eigenmodes of a cavity with astigmatic mirrors--they simply have different Gaussian beam parameters in the x and y axes, 
and consequently different Gouy phases, $\Psi_{x}(z)$ and $\Psi_{y}(z)$ (see Eqn.~\ref{equ:HGnm}). For an astigmatic beam, the  total phase lag of a HG mode when compared to a plane wave is then \cite{Bond2017}
\begin{equation}
\varphi=\left(n+\frac{1}{2}\right) \Psi_{x}(z)+\left(m+\frac{1}{2}\right) 
\Psi_{y}(z).
\end{equation}
This is in contrast to the LG modes, which are not separable in the x and y axes. 

Furthermore, the presence of astigmatism in the cavity eigenmode actually helps to break the degeneracy between resonances of HG modes of the same order. 
Consider the two extreme HG modes of order 6: the HG$_{60}$ mode has a phase lag $\varphi_{60}$ of $\left(6+\frac{1}{2}\right)\Psi_{x}(z)+\frac{1}{2} \Psi_{y}(z)$, therefore its resonance condition is dominated by the mirror curvatures along the x-axis. Meanwhile the HG$_{06}$ mode resonance condition depends primarily on the curvature along the y-axis since the phase lag $\varphi_{06}$ in this case is $\frac{1}{2}\Psi_{x}(z)+\left(6+\frac{1}{2}\right) \Psi_{y}(z)$. A large difference in x and y axis curvatures (i.e. astigmatism) therefore separates out the resonance conditions of these modes of the same order. 

\begin{figure}[htbp]
\centering
\includegraphics[width=\linewidth]{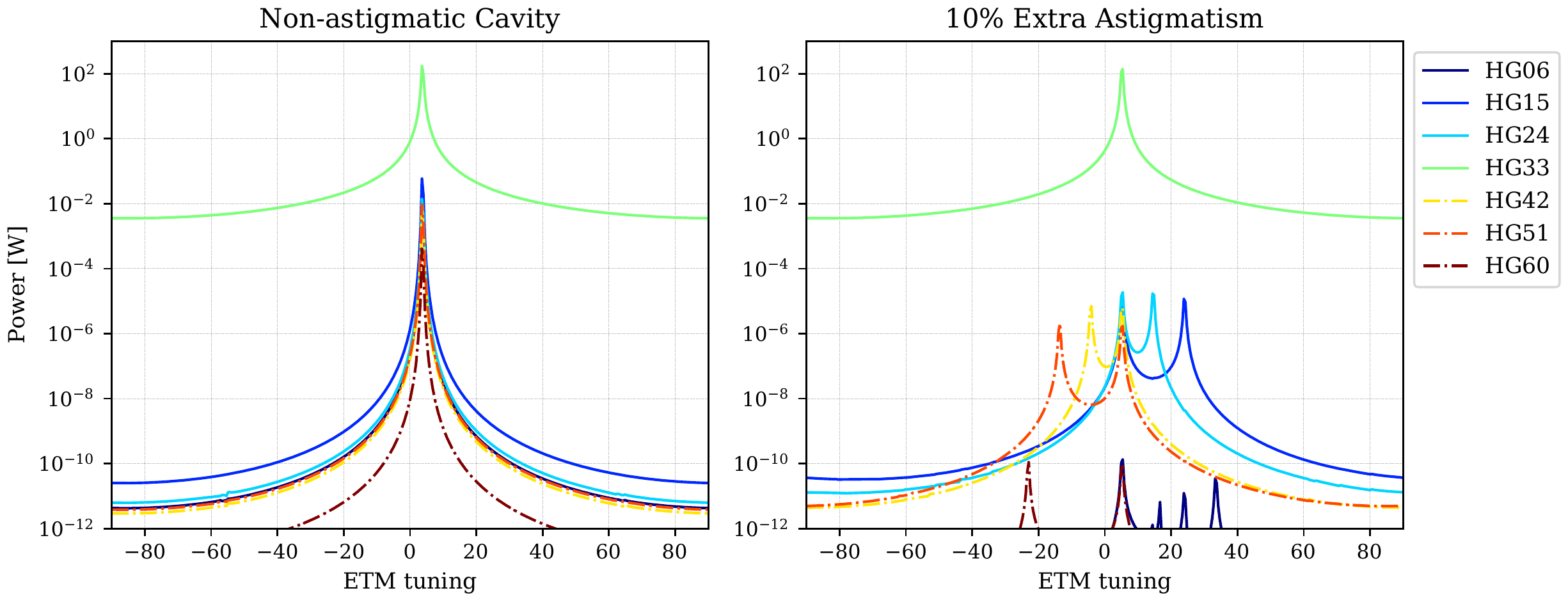}
\caption{Cavity circulating powers in HG modes of order 6 with map "ETM05\_S1\_Figure" applied to one cavity mirror. The left plot shows the co-resonance of the 6th order modes when no additional astigmatism is applied; the right plot has 400\,nm astigmatism added to the cavity mirrors. This causes the resonances of the 6th order modes to separate, leading to higher purity in the HG$_{33}$ mode.
}

\label{fig:HG33withZ22}
\end{figure}

In our case, adding sufficient astigmatism to separate out these resonances leads to a situation where coupling from the HG$_{33}$ mode to e.g. the HG$_{24}$ mode is effectively no worse for the mode purity, mode loss and contrast defect than coupling to modes of other orders. We can see the effects of astigmatism on the pseudo-degenerate high-order modes in Fig.~\ref{fig:HG33withZ22}. The left panel of Fig.~\ref{fig:HG33withZ22} shows the HG order 6 mode content inside a cavity with a random map applied to one mirror, as a function of cavity length tuning. 
The random map causes some coupling from the HG$_{33}$ mode into other HG modes of order 6. Since these modes are co-resonant, their circulating powers are quite large at the resonant tuning for the HG$_{33}$ mode. On the other hand, when we added 400\,nm ($\sim$10\,\%) astigmatism to the mirrors, as shown in the right panel, the resonances of the order 6 modes are separated out due to the unequal round-trip Gouy phases of the modes. Even though there is still coupling from the HG$_{33}$ mode into other order 6 modes, these modes are now non-resonant at the resonant tuning for the HG$_{33}$ mode. The curves shown for other HG modes of order 6 in the right panel of Fig.~\ref{fig:HG33withZ22} have two equal maxima: one at the resonant tuning for that specific mode, and one at the resonant tuning of the HG$_{33}$ mode which is the primary source of light scattered into the other modes of order 6. The relevant maximum for determining the interferometer performance is at the HG$_{33}$ resonant tuning, where the cavities will operate. In Fig.~\ref{fig:HG33withZ22} we can see that these maxima are orders of magnitude lower in the astigmatic case than the non-astigmatic case.
We therefore expect to see a large improvement in all figures of merit when increasing the astigmatism of the mirrors, even approaching the behavior of the HG$_{00}$ mode (which is already non-degenerate).

\textsc{Finesse} automatically calculates astigmatic cavity eigenmodes based on the radii of curvature $Rc_x$ and $Rc_y$, specified for the mirrors in x and y axes respectively. This is not the case however when astigmatism is specified for a mirror by inclusion in a mirror map. The advantage in our case of using the $Rc_x\neq Rc_y$ definition of astigmatic mirrors in \textsc{Finesse} is that it allows us to automatically mode match the input beam to the astigmatic eigenmode. As a result we are able to separate the effects of a changing cavity eigenmode with a fixed input mode (mode mismatch), from the more fundamental intra-cavity mode coupling effects in which we are primarily interested. To do this we need to calculate the extra curvature that should be added to the original mirror in terms of the Zernike coefficient $A_{2}^{2}$. 

The functional form of $Z_{2}^{2}$ with amplitude $A_{2}^{2}$ is
\begin{equation}
    Z_{2}^{2} = A_{2}^{2} \frac{\rho^{2}}{R_{m}^{2}} \cos 2 \theta = 
   \begin{cases}
   A_{2}^{2} \frac{\rho^{2}}{R_{m}^{2}}&\mbox{in the x axis} \\
   -A_{2}^{2} \frac{\rho^{2}}{R_{m}^{2}}&\mbox{in the y axis}
   \end{cases}
   \label{equ: curvature}
\end{equation}
where $R_{m}=0.15\,\rm{m}$ is the radius of the mirror.

This describes parabolic curves of opposite signs in the x and y axes. The equivalent spherical curvature, $K$ at $\rho=0$,  $\theta=0$ is given by
\begin{equation}
    K=\frac{\left|Z^{\prime \prime}(\rho = 0)\right|}{\left[1+\left(Z^{\prime}(\rho = 0)\right)^{2}\right]^{\frac{3}{2}}}.
\label{eqn:curv2}\end{equation}
Substituting the functional form in Eqn.~\ref{equ: curvature} into Eqn.~\ref{eqn:curv2}, we arrive at the equivalent spherical curvatures in the $x$ and $y$ axes:
\begin{equation}
    K = 
   \begin{cases}
   2\cdot A_{2}^{2} / R_{m}^2 &\mbox{in the x axis} \\
   -2\cdot A_{2}^{2} / R_{m}^2&\mbox{in the y axis},
   \end{cases}
\end{equation}
so the extra curvature is proportional to the Zernike coefficient, as expected. These curvatures can be added to the already present curvature (a.k.a. defocus) present in the mirror, and the sum inverted to find the equivalent radii of curvature.
Some examples of conversion from the Zernike coefficient $A_{2}^{2}$ to the relative change in radius of curvature are shown in Tab.~\ref{tab:radiusofcurvature}.

\begin{table}[htb]
\centering
\caption{ Zernike coefficients $A_{2}^{2}$ and their corresponding relative changes in radius of curvature ($Rc=2679.93\,{\rm m}$).}
\begin{tabular}{ccccc}
&$A_{2}^{2}$ [nm] &($Rc_{y}$-$Rc$)/$Rc$ & ($Rc_{x}$-$Rc$)/$Rc$  \\
\hline\hline
0&10&	0.002388 & -0.002388\\
1&30&  0.007199 & -0.007199\\
2&50&	0.012056 & -0.012056\\
3&80&  0.019430 & -0.019430\\
4&100&	0.024406 & -0.024406\\
5&400&  0.105337 & -0.105337\\
\hline\hline
\end{tabular}
  \label{tab:radiusofcurvature}
\end{table} 

\begin{figure}[tbp]
\centering
\includegraphics[width=\linewidth]{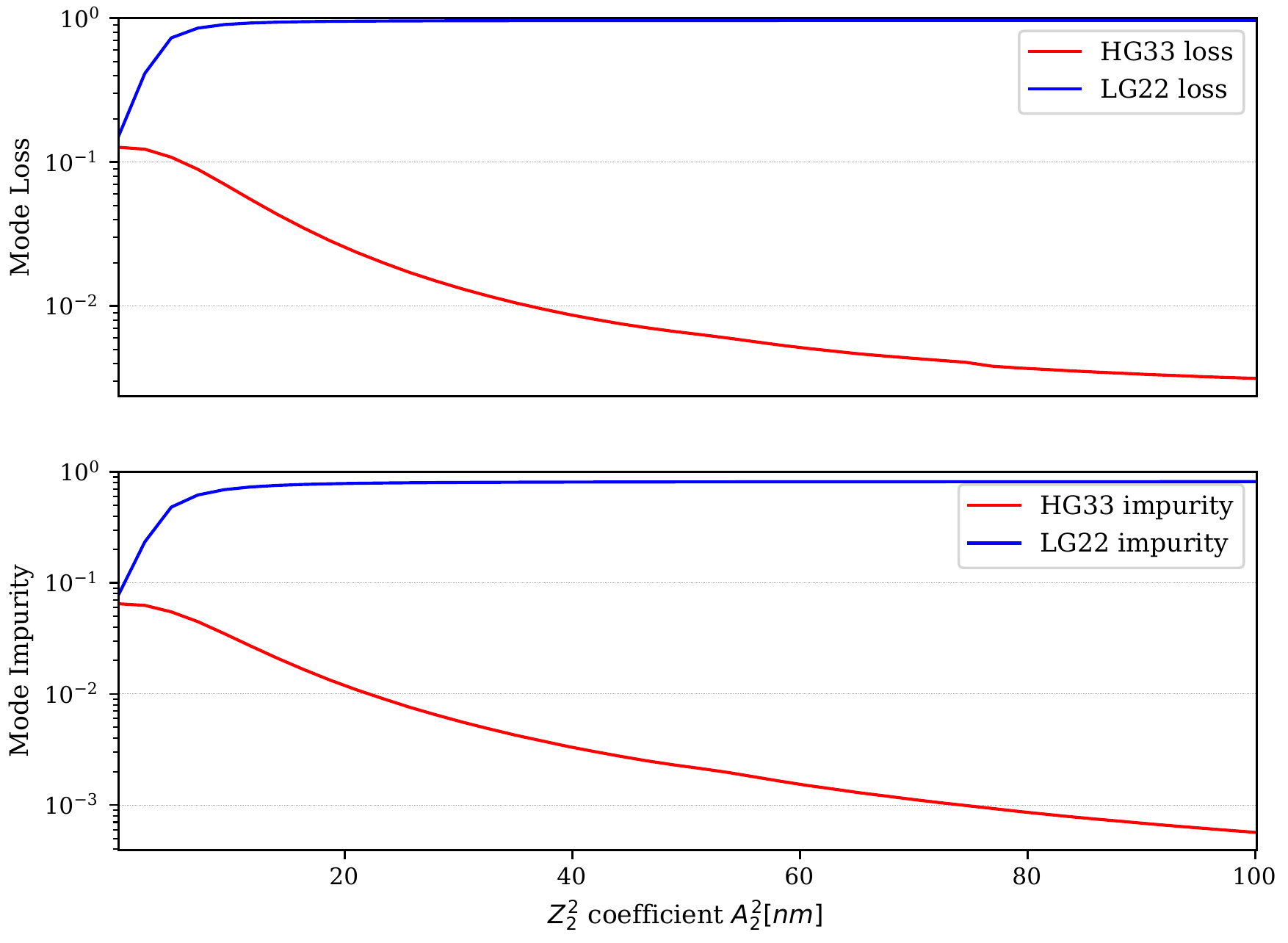}
\caption{Mode purity and losses of the HG$_{33}$ and LG$_{22}$ modes with increasingly astigmatic cavity eigenmodes.}
\label{fig:lossandpurityastigmatism}
\end{figure}

The original map file was decomposed into the Zernike basis according to Eqn.~\ref{equ:zernikecoeff} giving a $Z_{2}^{2}$ amplitude, corresponding to vertical astigmatism, of 0.12\,nm. We gradually increased the astigmatism by 100\,nm by adjusting the cavity mirror radii of curvature differentially in the x and y axes. The mode purity and losses were calculated for both LG$_{22}$ and HG$_{33}$ modes, and the results are shown in Fig.~\ref{fig:lossandpurityastigmatism}.

\begin{figure}[htbp]
\centering
\includegraphics[width=\linewidth]{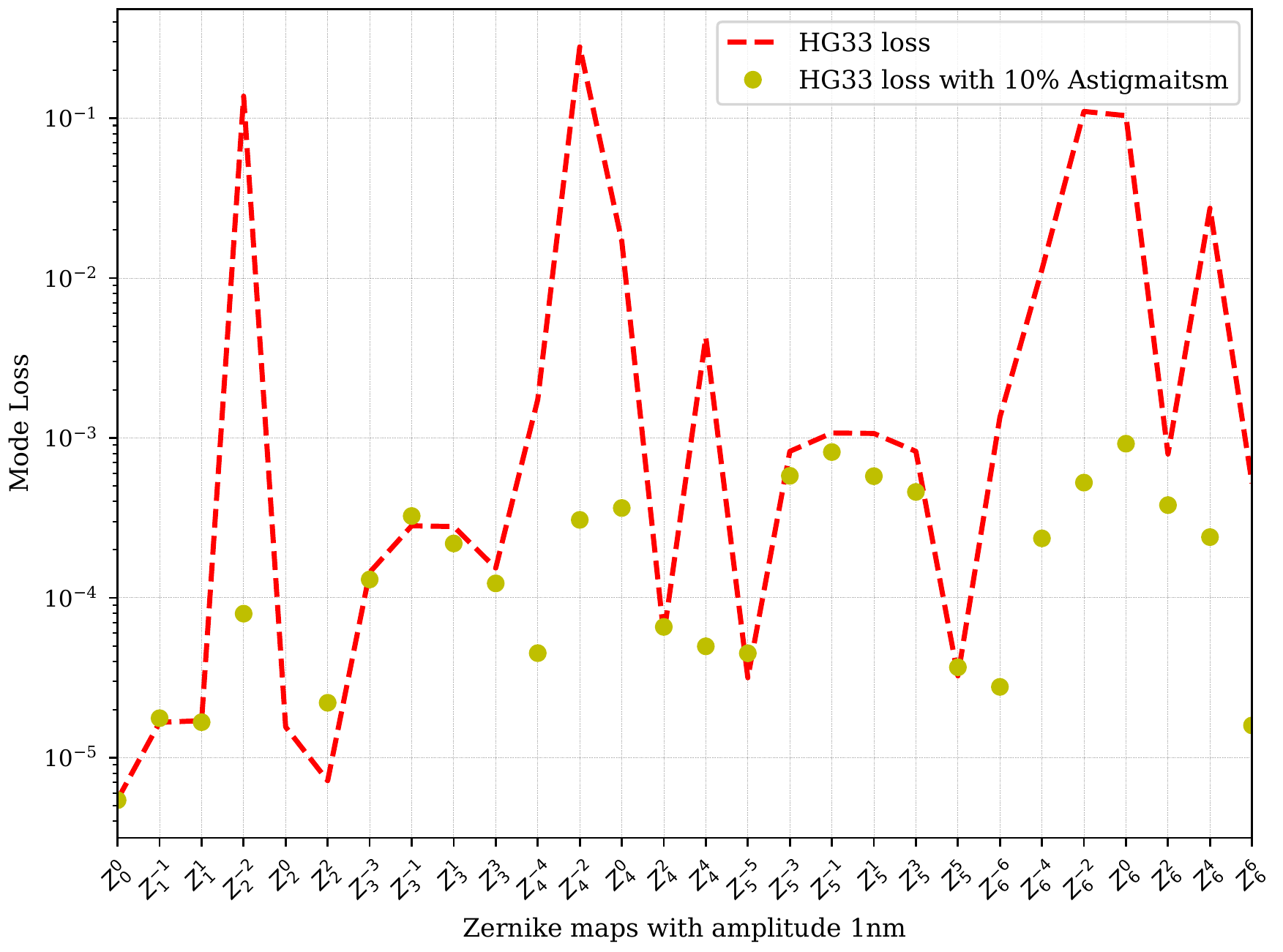}
\caption{The mode loss and impurity of the HG33 mode for each individual Zernike term with amplitude 1\,nm. The red line shows the same result in Fig.~\ref{fig:modelossandpurityLGtwHGt}. The yellow dots show the same configuration but with 400\,nm extra astigmatism added.}
\label{fig:zernikeastigmatism}
\end{figure}

Fig.~\ref{fig:lossandpurityastigmatism} shows that as we increase the astigmatism of the mirrors the HG$_{33}$ mode loss and impurity decrease, in contrast to the LG$_{22}$ mode which shows rapidly increasing loss and impurity with increasing astigmatism. We can also see the effect of astigmatism on the HG$_{33}$ mode performance by looking at the mode loss for individual Zernike terms, but this time with the extra astigmatism applied. Fig.~\ref{fig:zernikeastigmatism} compares losses with individual Zernike terms for the HG$_{33}$ mode in the original case, and the case with $\sim$400\,nm astigmatism applied by modifying $Rc_{x}$ and $Rc_{y}$ as previously described~\footnote{400\,nm astigmatism corresponds to about 10\% change of the radius of curvature - see Tab.~\ref{tab:radiusofcurvature}.}. Here we see that the effects of different Zernike terms are more similar to each other, and in general lower, for the astigmatic case than the original case. This is understood to be because the particularly problematic Zernike terms in the original case (e.g. $Z_{2}^{-2}$ and $Z_{4}^{-2}$) caused strong coupling to other HG modes of order 6, which were co-resonant with the HG$_{33}$ mode. In the astigmatic case these HG modes are no longer co-resonant, and so the previously problematic Zernike terms have an impact similar to any other Zernike terms.

We can utilize this result and add 400\,nm astigmatism to the test masses before rotation to further improve the performance of HG$_{33}$ mode in the simulated interferometer. The optimal rotation angles now should be
\begin{equation}
    \phi = \arctan(A_{2}^{-2}/(A_{2}^{2}+\Delta A))/2
    \label{equ:rotationangle}
\end{equation}
where $\Delta A$ represents the extra astigmatism added. The optimal rotation angles of 
five random maps with 400\,nm of astigmatism added are shown in Tab.~\ref{tab:rotationangle} -- the required angles are now much smaller than the non-astigmatic case. The radii of curvature of the test masses are now different from Tab.~\ref{tab:cavityparameters} since we are implementing astigmatism by setting different radii of curvature in the x and y axes, as shown in Tab.~\ref{tab:newcavity}. It was also necessary to add extra defocus to the mirrors to keep the clipping loss at 1\,ppm for the astigmatic beams. The coating thermal noise calculated from the fluctuation-dissipation theorem\cite{PhysRevD.82.042003} with astigmatic beam scales as $1/(w_{x} \cdot w_{y})$ instead of $1/w^2$ as in the non-astigmatic case. For the beam sizes considered here the coating thermal noise PSD will increase by only 1\% when the astigmatism is 10~\%.

\begin{table}[htbp]
\centering
\caption{Rotation angles used by five random maps in the case of no added astigmatism, $\phi_{0}$, and with 400\,nm added astigmatism, $\phi_{400nm}$.
}
\begin{tabular}{ccc}
map ID&$\phi_{0}$[deg]&$\phi_{400\mathrm{nm}}$[deg]   \\
\hline\hline
0&28.9&$1.23\cdot 10^{-2}$\\
1&-2.4&$-1.18\cdot 10^{-3}$\\
2&27.1&$-1.17\cdot 10^{-2}$\\
3&-30.5&$-1.27\cdot 10^{-2}$\\
4&40.1&$1.42\cdot 10^{-2}$\\
\hline\hline
\end{tabular}
  \label{tab:rotationangle}
\end{table}

\begin{table}[tbp]
\centering
\caption{Radii of curvature of the arm cavity mirrors for HG$_{33}$ corresponding to 400\,nm additional astigmatism plus additional defocus to keep the clipping loss at 1\,ppm.}
\begin{tabular}{cc|cc}
\multicolumn{2}{c}{ITM} &\multicolumn{2}{c}{ETM}\\
\hline \hline 
$\mathrm{R}_{cx}$ [m]  & $\mathrm{R}_{cy}$ [m] & $\mathrm{R}_{cx}$ [m]  & $\mathrm{R}_{cy}$ [m] \\
-2516.89& -3109.47 & 2516.89 & 3109.47 \\
\hline \hline 
\end{tabular}
\label{tab:newcavity}

\end{table}

The result for the loss and impurity are shown in green in Figs.~\ref{fig:rotationloss} and~\ref{fig:rotationpurity} respectively. The losses and impurity are significantly reduced from even the rotated case. The average and standard deviation of the loss for HG$_{33}$ mode with the rotated maps after adding 400\,nm of astigmatism is 82.9\,ppm and 16.4\,ppm respectively. And for the impurity, the average and standard deviation are 5.3ppm and 6.3ppm respectively. These figures of merit are now close to the loss and impurity for HG$_{00}$ case. The loss has been reduced by more than two orders of magnitude by adding 400\,nm of astigmatism and rotating the maps to minimize the $Z_{2}^{-2}$ term. The impurity on the other hand has been reduced by about three orders of magnitude. In the astigmatic case the HG$_{33}$ mode losses due to surface figure errors are well below the typical aLIGO arm cavity losses of 12000\,ppm (shown as the dashed vertical line in Fig.~\ref{fig:rotationloss}), which is equivalent to 85\,ppm loss per cavity round trip. This is an important step towards showing their compatibility with aLIGO-like interferometers. It should be noted that in this simulation we only consider the loss and contrast defect caused by the low frequency mirror distortions and ignore other known or unknown factors, such as wide angle scatter from mirror surface roughness and coating absorption, that also contribute in real aLIGO experiments. This explains why the currently used HG$_{00}$ mode distribution does not center at the typical aLIGO measured values in Fig.~\ref{fig:histograms}. 


The contrast defect of HG$_{33}$ has also been greatly reduced by adding astigmatism and rotating the mirrors, as shown in Fig.~\ref{fig:cd10nm}. The average and standard deviation of the contrast defect for HG$_{33}$ mode with the rotated maps after adding 400\,nm of astigmatism is 10.5\,ppm and 12.9\,ppm respectively. It is again much closer to the contrast defect for the HG$_{00}$ case, and well below the typical aLIGO measured contrast defect of 400\,ppm. Adding 400\,nm of astigmatism and rotating the maps has reduced the contrast defect by more than three orders of magnitude. Once again this shows that the impact of test mass surface figure errors on the HG$_{33}$ mode performance can be made negligible by deliberately adding astigmatism to the surface figures.

\section{Conclusions}
\label{sec:conclusion}

We have investigated the performance of the HG$_{33}$ and LG$_{22}$ modes against surface deformations representative of those expected in next-generation aLIGO test mass mirrors. Simulations were performed to assess the performance of these modes in aLIGO-like arm cavities and a Fabry–Perot Michelson interferometer. This investigation has demonstrated that without mirror modifications, higher-order Hermite-Gauss modes are only marginally more robust against figure errors than Laguerre-Gauss modes of the same order. However with the deliberate addition of vertical astigmatism to the mirrors, we found that the HG$_{33}$ mode performs almost as well as the HG$_{00}$ mode in terms of the metrics of arm loss, mode purity and contrast defect considered here. The loss and contrast defect of the HG$_{33}$ mode, with the addition of astigmatism, were also seen to be well below the typical aLIGO measured values. This indicates that the effects of mirror surface flatness errors will not be a limiting factor for this mode. 

There remain some aspects of future gravitational-wave detector performance with higher-order HG modes that still require further study however. This includes HG mode generation at high powers and purities~\cite{Ast}, squeezing of higher-order HG modes~\cite{Heinze}, alignment and mode matching requirements~\cite{Jones}, alignment and mode matching sensing and control, and susceptibility to parametric instabilities. Nonetheless, we have demonstrated here that one of the main problems associated with higher-order LG modes for future gravitational wave detectors--fragility against mirror surface figure imperfections--can be effectively sidestepped for HG modes by using astigmatic cavity mirrors. 

\section{Acknowledgments}
We thank GariLynn Billingsley for helpful discussions about Advanced LIGO mirror maps. This work was supported by National Science Foundation grants PHY-1806461 and PHY-2012021. 

\appendix
\section{Random map generation}
\label{sec:randommaps}
We use the Zernike basis to describe the low spatial frequency distortion of a mirror surface and map decomposition and reconstruction. Zernike polynomials are a complete
set of functions which are orthogonal over the unit disc and defined by radial index, n, and azimuthal index, m, with $m \leq n$. For any index m we have~\cite{Bond2017}
\begin{equation}
\begin{aligned}
&Z_{n}^{+m}(\rho, \phi)=\cos (m \phi) R_{n}^{m}(\rho) \quad \text {the even polynomial}\\
&Z_{n}^{-m}(\rho, \phi)=\sin (m \phi) R_{n}^{m}(\rho) \quad \text {the odd polynomial}
\end{aligned}
\end{equation}
with $\rho$ the normalised radius, $\phi$ the azimuthal angle and $R_{n}^{m}(\rho)$ the radial function
\begin{equation}
R_{n}^{m}(\rho)=\left\{\begin{array}{ll}
\sum_{h=0}^{\frac{1}{2}(n-m)} \frac{(-1)^{h}(n-h) !}{h !\left(\frac{1}{2}(n+m)-h\right) !\left(\frac{1}{2}(n-m)-h\right) !} \rho^{n-2 h} & \text {n-m is even}  \\
0 & \text {n-m is odd}
\end{array}\right.
\end{equation}

Generation of the random maps was achieved by first decomposing a measured aLIGO map into the Zernike basis and obtaining the Zernike coefficients $A_{n}^{m}$ using the following formula:
\begin{equation}
A_{n}^{m}=
\frac{\sum_{x, y} Z_{\operatorname{map}} (x, y) \cdot Z_{n}^{m}(x, y)}{\sum_{x, y} Z_{n}^{m}(x, y) \cdot Z_{n}^{m}(x, y)},
\label{equ:zernikecoeff}
\end{equation}
where $Z_{\operatorname{map}}$ represents the surface deformations of the mirror, and $Z_{n}^{m}$ is the Zernike polynomial with radial index n and azimuthal index m. 

\begin{figure}[htbp]
\centering
\includegraphics[width=\linewidth]{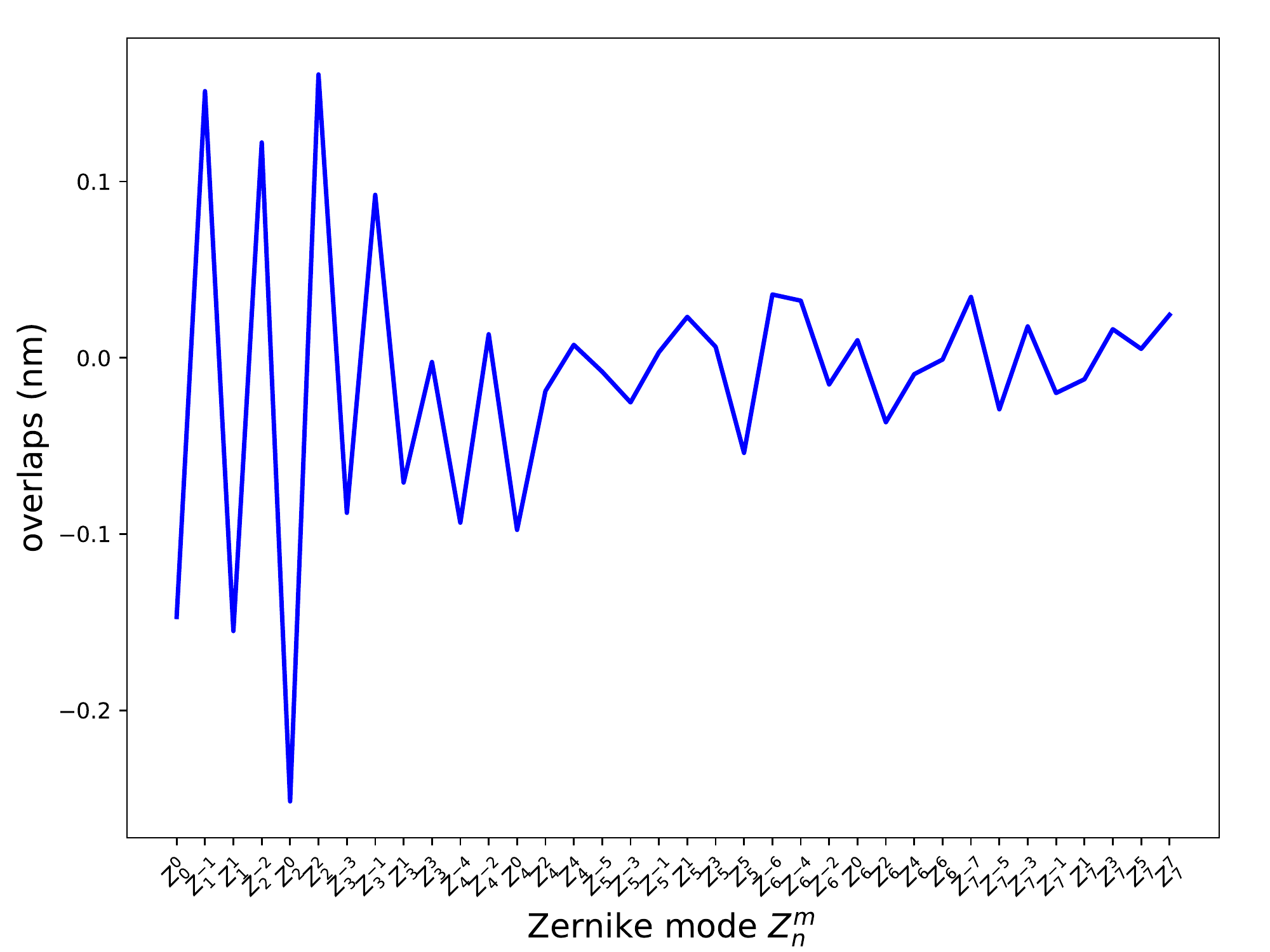}
\caption{Decomposition of mirror surface map "ETM\_05\_S1\_Figure" into the Zernike basis}
\label{fig:decomposition}
\end{figure}

\begin{figure}[htbp]
\centering
\includegraphics[width=\linewidth]{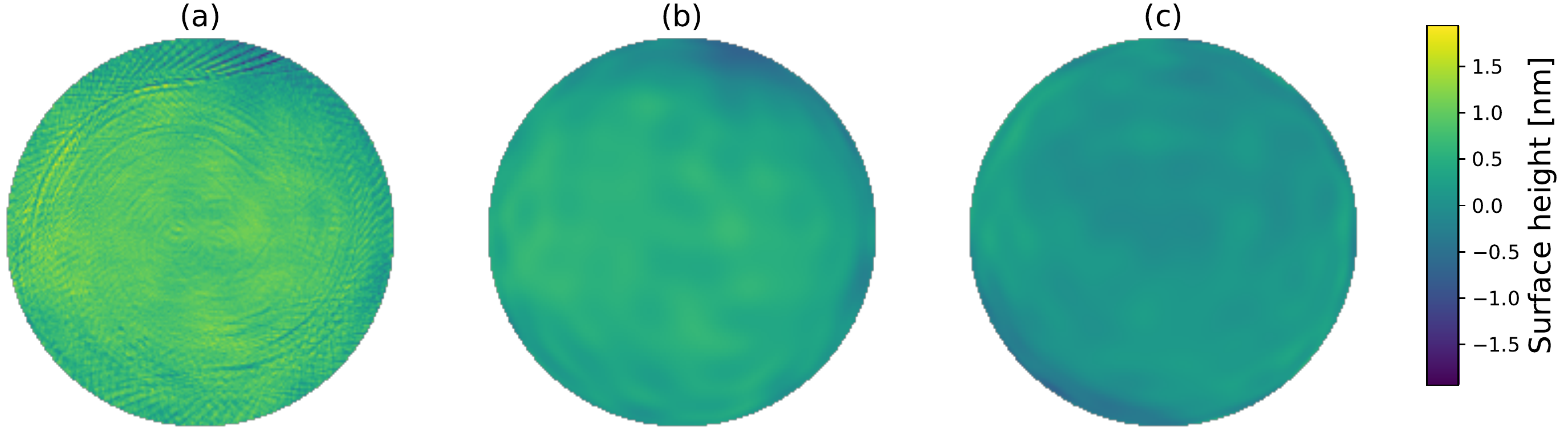}
\caption{Example mirror maps: (a) the original `ETM\_05\_Figure' map, (b) the `ETM\_05\_Figure' map recomposed from Zernike polynomials up to radial index 25, and (c) a random mirror map, generated to have a similar spatial frequency spectrum to (b).}
\label{fig:randommap}
\end{figure}

Since all random maps will have the same spatial frequency characteristics as the measured map which was initially decomposed in the Zernike basis, the choice of this `base' map was important. The surface figures of current aLIGO mirrors are understood to deviate from their intended figures primarily due to non-uniformities in the applied high-reflective coatings~\cite{LMAcoatings}. A procedure involving careful analysis of coating non-uniformities and pre-emptive polishing to cancel them out is expected to lead to the next generation of aLIGO mirrors having coated surface figures which are roughly equivalent to the uncoated figures of current aLIGO maps. Looking forward to next generation of mirrors, we decided therefore to use an uncoated aLIGO mirror map "ETM\_05\_S1\_Figure" as the base map for random map generation. Fig.~\ref{fig:decomposition} shows the results of the Zernike decomposition of this map, while the map itself is shown in the leftmost panel of Fig.~\ref{fig:randommap}.

After Zernike decomposition we then define the `layer coefficients'
$B_n$ as the quadrature sums of coefficients for all Zernike terms with the same radial index $n$:
\begin{equation}
B_n=\sqrt{\sum_{m}\left(A_{n}^{m}\right)^{2}}.
\end{equation}
The $B_{0}, B_{1}$ layer coefficients (representing piston, pitch and yaw), and $A_2^0$ (representing defocus) are set to zero since these are degrees of freedom which can be controlled actively in an interferometer. 
Zernike terms with large $n$ are expected to cause wide-angle scatter, little of which will be into pseudo-degenerate modes of order 6. Therefore these high spatial frequency surface features are expected to contribute minimally to the pseudo-degeneracy problem that is the primary focus of the work reported here. Layer coefficients in the simulation are calculated up to $n=25$, which is several layers beyond the point at which the simulation results were observed to converge, matching the expected behaviour.

In order to generate random maps with the same layer coefficients as above, we randomly redistributed the layer coefficients between the Zernike coefficients within that layer, and combined all the Zernike coefficients in the Zernike basis to formulate the random maps. The new random Zernike coefficients are
\begin{equation}
{A^{\prime}}_{n}^{m} = B_{n} \frac{\Lambda_{n}^{m}}{\sqrt{\sum_{m}\left(\Lambda_{n}^{m}\right)^{2}}},
\end{equation}
where $\Lambda_{n}^{m}$ are random numbers taken from a uniform distribution from -0.5 to 0.5, i.e. $\Lambda_{n}^{m} \sim U(-0.5, 0.5)$. ${A^{\prime}}_{n}^{m}$ is normalized in this way such that the layer coefficients calculated from these new Zernike coefficients ${A^{\prime}}_{n}^{m}$ are the same as $B_{n}$, which are the layer coefficients from the aLIGO-measured map ``ETM\_05\_S1\_Figure". Then the random map $\mathrm{Z^{Rand}_{map}}$ can be constructed by recombining the Zernike polynomials $Z_{n}^{m}$ with the new random Zernike coefficients ${A^{\prime}}_{n}^{m}$ we obtained earlier
\begin{equation}
\mathrm{Z^{Rand}_{map} = \sum_{n} \sum_{m} {A^{\prime}}_{n}^{m} \cdot Z_{n}^{m}}.
\end{equation}

The random maps generated this way will have roughly the same spatial frequency spectra as the aLIGO measured map used, as demonstrated in Ref.~\cite{Bond2017}. Fig.~\ref{fig:randommap} shows one of the example random maps alongside the original aLIGO base map, as well as the base map recomposed from Zernike polynomials with radial index up to 25


\bibliographystyle{unsrt}
\bibliography{template}

\begin{thebibliography}{10}

\bibitem{aLIGO}
J~Aasi, B~P Abbott, R~Abbott, T~Abbott, M~R Abernathy, K~Ackley, C~Adams,
  T~Adams, P~Addesso, R~X Adhikari, V~Adya, C~Affeldt, N~Aggarwal, O~D Aguiar,
  A~Ain, P~Ajith, A~Alemic, B~Allen, D~Amariutei, S~B Anderson, W~G Anderson,
  K~Arai, M~C Araya, C~Arceneaux, J~S Areeda, G~Ashton, S~Ast, S~M Aston,
  P~Aufmuth, C~Aulbert, B~E Aylott, S~Babak, P~T Baker, S~W Ballmer, J~C
  Barayoga, M~Barbet, S~Barclay, B~C Barish, D~Barker, B~Barr, L~Barsotti,
  J~Bartlett, M~A Barton, I~Bartos, R~Bassiri, J~C Batch, C~Baune, B~Behnke,
  A~S Bell, C~Bell, M~Benacquista, J~Bergman, G~Bergmann, C~P~L Berry,
  J~Betzwieser, S~Bhagwat, R~Bhandare, I~A Bilenko, G~Billingsley, J~Birch,
  S~Biscans, C~Biwer, J~K Blackburn, L~Blackburn, C~D Blair, D~Blair, O~Bock,
  T~P Bodiya, P~Bojtos, C~Bond, R~Bork, M~Born, Sukanta Bose, P~R Brady, V~B
  Braginsky, J~E Brau, D~O Bridges, M~Brinkmann, A~F Brooks, D~A Brown, D~D
  Brown, N~M Brown, S~Buchman, A~Buikema, A~Buonanno, L~Cadonati,
  J~Calder{\'{o}}n Bustillo, J~B Camp, K~C Cannon, J~Cao, C~D Capano, S~Caride,
  S~Caudill, M~Cavagli{\`{a}}, C~Cepeda, R~Chakraborty, T~Chalermsongsak, S~J
  Chamberlin, S~Chao, P~Charlton, Y~Chen, H~S Cho, M~Cho, J~H Chow,
  N~Christensen, Q~Chu, S~Chung, G~Ciani, F~Clara, J~A Clark, C~Collette,
  L~Cominsky, M~Constancio, D~Cook, T~R Corbitt, N~Cornish, A~Corsi, C~A Costa,
  M~W Coughlin, S~Countryman, P~Couvares, D~M Coward, M~J Cowart, D~C Coyne,
  R~Coyne, K~Craig, J~D~E Creighton, T~D Creighton, J~Cripe, S~G Crowder,
  A~Cumming, L~Cunningham, C~Cutler, K~Dahl, T~Dal Canton, M~Damjanic, S~L
  Danilishin, K~Danzmann, L~Dartez, I~Dave, H~Daveloza, G~S Davies, E~J Daw,
  D~DeBra, W~Del Pozzo, T~Denker, T~Dent, V~Dergachev, R~T DeRosa, R~DeSalvo,
  S~Dhurandhar, M~Díaz, I~Di Palma, G~Dojcinoski, E~Dominguez, F~Donovan, K~L
  Dooley, S~Doravari, R~Douglas, T~P Downes, J~C Driggers, Z~Du, S~Dwyer,
  T~Eberle, T~Edo, M~Edwards, M~Edwards, A~Effler, H.-B Eggenstein, P~Ehrens,
  J~Eichholz, S~S Eikenberry, R~Essick, T~Etzel, M~Evans, T~Evans,
  M~Factourovich, S~Fairhurst, X~Fan, Q~Fang, B~Farr, W~M Farr, M~Favata,
  M~Fays, H~Fehrmann, M~M Fejer, D~Feldbaum, E~C Ferreira, R~P Fisher, Z~Frei,
  A~Freise, R~Frey, T~T Fricke, P~Fritschel, V~V Frolov, S~Fuentes-Tapia,
  P~Fulda, M~Fyffe, J~R Gair, S~Gaonkar, N~Gehrels, L~{\'{A}} Gergely, J~A
  Giaime, K~D Giardina, J~Gleason, E~Goetz, R~Goetz, L~Gondan,
  G~Gonz{\'{a}}lez, N~Gordon, M~L Gorodetsky, S~Gossan, S~Go{\ss}ler, C~Gräf,
  P~B Graff, A~Grant, S~Gras, C~Gray, R~J~S Greenhalgh, A~M Gretarsson,
  H~Grote, S~Grunewald, C~J Guido, X~Guo, K~Gushwa, E~K Gustafson, R~Gustafson,
  J~Hacker, E~D Hall, G~Hammond, M~Hanke, J~Hanks, C~Hanna, M~D Hannam,
  J~Hanson, T~Hardwick, G~M Harry, I~W Harry, M~Hart, M~T Hartman, C-J Haster,
  K~Haughian, S~Hee, M~Heintze, G~Heinzel, M~Hendry, I~S Heng, A~W Heptonstall,
  M~Heurs, M~Hewitson, S~Hild, D~Hoak, K~A Hodge, S~E Hollitt, K~Holt,
  P~Hopkins, D~J Hosken, J~Hough, E~Houston, E~J Howell, Y~M Hu, E~Huerta,
  B~Hughey, S~Husa, S~H Huttner, M~Huynh, T~Huynh-Dinh, A~Idrisy, N~Indik, D~R
  Ingram, R~Inta, G~Islas, J~C Isler, T~Isogai, B~R Iyer, K~Izumi, M~Jacobson,
  H~Jang, S~Jawahar, Y~Ji, F~Jim{\'{e}}nez-Forteza, W~W Johnson, D~I Jones,
  R~Jones, L~Ju, K~Haris, V~Kalogera, S~Kandhasamy, G~Kang, J~B Kanner,
  E~Katsavounidis, W~Katzman, H~Kaufer, S~Kaufer, T~Kaur, K~Kawabe, F~Kawazoe,
  G~M Keiser, D~Keitel, D~B Kelley, W~Kells, D~G Keppel, J~S Key,
  A~Khalaidovski, F~Y Khalili, E~A Khazanov, C~Kim, K~Kim, N~G Kim, N~Kim, Y.-M
  Kim, E~J King, P~J King, D~L Kinzel, J~S Kissel, S~Klimenko, J~Kline,
  S~Koehlenbeck, K~Kokeyama, V~Kondrashov, M~Korobko, W~Z Korth, D~B Kozak,
  V~Kringel, B~Krishnan, C~Krueger, G~Kuehn, A~Kumar, P~Kumar, L~Kuo, M~Landry,
  B~Lantz, S~Larson, P~D Lasky, A~Lazzarini, C~Lazzaro, J~Le, P~Leaci,
  S~Leavey, E~O Lebigot, C~H Lee, H~K Lee, H~M Lee, J~R Leong, Y~Levin,
  B~Levine, J~Lewis, T~G~F Li, K~Libbrecht, A~Libson, A~C Lin, T~B Littenberg,
  N~A Lockerbie, V~Lockett, J~Logue, A~L Lombardi, M~Lormand, J~Lough, M~J
  Lubinski, H~Lück, A~P Lundgren, R~Lynch, Y~Ma, J~Macarthur, T~MacDonald,
  B~Machenschalk, M~MacInnis, D~M Macleod, F~Maga{\~{n}}a-Sandoval, R~Magee,
  M~Mageswaran, C~Maglione, K~Mailand, I~Mandel, V~Mandic, V~Mangano, G~L
  Mansell, S~M{\'{a}}rka, Z~M{\'{a}}rka, A~Markosyan, E~Maros, I~W Martin, R~M
  Martin, D~Martynov, J~N Marx, K~Mason, T~J Massinger, F~Matichard, L~Matone,
  N~Mavalvala, N~Mazumder, G~Mazzolo, R~McCarthy, D~E McClelland, S~McCormick,
  S~C McGuire, G~McIntyre, J~McIver, K~McLin, S~McWilliams, G~D Meadors,
  M~Meinders, A~Melatos, G~Mendell, R~A Mercer, S~Meshkov, C~Messenger, P~M
  Meyers, H~Miao, H~Middleton, E~E Mikhailov, A~Miller, J~Miller, M~Millhouse,
  J~Ming, S~Mirshekari, C~Mishra, S~Mitra, V~P Mitrofanov, G~Mitselmakher,
  R~Mittleman, B~Moe, S~D Mohanty, S~R~P Mohapatra, B~Moore, D~Moraru,
  G~Moreno, S~R Morriss, K~Mossavi, C~M Mow-Lowry, C~L Mueller, G~Mueller,
  S~Mukherjee, A~Mullavey, J~Munch, D~Murphy, P~G Murray, A~Mytidis, T~Nash,
  R~K Nayak, V~Necula, K~Nedkova, G~Newton, T~Nguyen, A~B Nielsen, S~Nissanke,
  A~H Nitz, D~Nolting, M~E~N Normandin, L~K Nuttall, E~Ochsner, J~O'Dell,
  E~Oelker, G~H Ogin, J~J Oh, S~H Oh, F~Ohme, P~Oppermann, R~Oram, B~O'Reilly,
  W~Ortega, R~O'Shaughnessy, C~Osthelder, C~D Ott, D~J Ottaway, R~S Ottens,
  H~Overmier, B~J Owen, C~Padilla, A~Pai, S~Pai, O~Palashov, A~Pal-Singh,
  H~Pan, C~Pankow, F~Pannarale, B~C Pant, M~A Papa, H~Paris, Z~Patrick,
  M~Pedraza, L~Pekowsky, A~Pele, S~Penn, A~Perreca, M~Phelps, V~Pierro, I~M
  Pinto, M~Pitkin, J~Poeld, A~Post, A~Poteomkin, J~Powell, J~Prasad, V~Predoi,
  S~Premachandra, T~Prestegard, L~R Price, M~Principe, S~Privitera, R~Prix,
  L~Prokhorov, O~Puncken, M~Pürrer, J~Qin, V~Quetschke, E~Quintero, G~Quiroga,
  R~Quitzow-James, F~J Raab, D~S Rabeling, H~Radkins, P~Raffai, S~Raja,
  G~Rajalakshmi, M~Rakhmanov, K~Ramirez, V~Raymond, C~M Reed, S~Reid, D~H
  Reitze, O~Reula, K~Riles, N~A Robertson, R~Robie, J~G Rollins, V~Roma, J~D
  Romano, G~Romanov, J~H Romie, S~Rowan, A~Rüdiger, K~Ryan, S~Sachdev,
  T~Sadecki, L~Sadeghian, M~Saleem, F~Salemi, L~Sammut, V~Sandberg, J~R
  Sanders, V~Sannibale, I~Santiago-Prieto, B~S Sathyaprakash, P~R Saulson,
  R~Savage, A~Sawadsky, J~Scheuer, R~Schilling, P~Schmidt, R~Schnabel, R~M~S
  Schofield, E~Schreiber, D~Schuette, B~F Schutz, J~Scott, S~M Scott,
  D~Sellers, A~S Sengupta, A~Sergeev, G~Serna, A~Sevigny, D~A Shaddock, M~S
  Shahriar, M~Shaltev, Z~Shao, B~Shapiro, P~Shawhan, D~H Shoemaker, T~L Sidery,
  X~Siemens, D~Sigg, A~D Silva, D~Simakov, A~Singer, L~Singer, R~Singh, A~M
  Sintes, B~J~J Slagmolen, J~R Smith, M~R Smith, R~J~E Smith, N~D
  Smith-Lefebvre, E~J Son, B~Sorazu, T~Souradeep, A~Staley, J~Stebbins,
  M~Steinke, J~Steinlechner, S~Steinlechner, D~Steinmeyer, B~C Stephens,
  S~Steplewski, S~Stevenson, R~Stone, K~A Strain, S~Strigin, R~Sturani, A~L
  Stuver, T~Z Summerscales, P~J Sutton, M~Szczepanczyk, G~Szeifert, D~Talukder,
  D~B Tanner, M~T{\'{a}}pai, S~P Tarabrin, A~Taracchini, R~Taylor, G~Tellez,
  T~Theeg, M~P Thirugnanasambandam, M~Thomas, P~Thomas, K~A Thorne, K~S Thorne,
  E~Thrane, V~Tiwari, C~Tomlinson, C~V Torres, C~I Torrie, G~Traylor, M~Tse,
  D~Tshilumba, D~Ugolini, C~S Unnikrishnan, A~L Urban, S~A Usman, H~Vahlbruch,
  G~Vajente, G~Valdes, M~Vallisneri, A~A van Veggel, S~Vass, R~Vaulin,
  A~Vecchio, J~Veitch, P~J Veitch, K~Venkateswara, R~Vincent-Finley, S~Vitale,
  T~Vo, C~Vorvick, W~D Vousden, S~P Vyatchanin, A~R Wade, L~Wade, M~Wade,
  M~Walker, L~Wallace, S~Walsh, H~Wang, M~Wang, X~Wang, R~L Ward, J~Warner,
  M~Was, B~Weaver, M~Weinert, A~J Weinstein, R~Weiss, T~Welborn, L~Wen,
  P~Wessels, T~Westphal, K~Wette, J~T Whelan, S~E Whitcomb, D~J White, B~F
  Whiting, C~Wilkinson, L~Williams, R~Williams, A~R Williamson, J~L Willis,
  B~Willke, M~Wimmer, W~Winkler, C~C Wipf, H~Wittel, G~Woan, J~Worden, S~Xie,
  J~Yablon, I~Yakushin, W~Yam, H~Yamamoto, C~C Yancey, Q~Yang, M~Zanolin, Fan
  Zhang, L~Zhang, M~Zhang, Y~Zhang, C~Zhao, M~Zhou, X~J Zhu, M~E Zucker,
  S~Zuraw, and J~Zweizig.
\newblock Advanced {LIGO}.
\newblock {\em Classical and Quantum Gravity}, 32(7):074001, mar 2015.

\bibitem{AdVirgo}
F~Acernese, M~Agathos, K~Agatsuma, D~Aisa, N~Allemandou, A~Allocca, J~Amarni,
  P~Astone, G~Balestri, G~Ballardin, F~Barone, J-P Baronick, M~Barsuglia,
  A~Basti, F~Basti, Th~S Bauer, V~Bavigadda, M~Bejger, M~G Beker, C~Belczynski,
  D~Bersanetti, A~Bertolini, M~Bitossi, M~A Bizouard, S~Bloemen, M~Blom,
  M~Boer, G~Bogaert, D~Bondi, F~Bondu, L~Bonelli, R~Bonnand, V~Boschi, L~Bosi,
  T~Bouedo, C~Bradaschia, M~Branchesi, T~Briant, A~Brillet, V~Brisson, T~Bulik,
  H~J Bulten, D~Buskulic, C~Buy, G~Cagnoli, E~Calloni, C~Campeggi, B~Canuel,
  F~Carbognani, F~Cavalier, R~Cavalieri, G~Cella, E~Cesarini,
  E~Chassande-Mottin, A~Chincarini, A~Chiummo, S~Chua, F~Cleva, E~Coccia, P-F
  Cohadon, A~Colla, M~Colombini, A~Conte, J-P Coulon, E~Cuoco, A~Dalmaz,
  S~D'Antonio, V~Dattilo, M~Davier, R~Day, G~Debreczeni, J~Degallaix,
  S~Del{\'{e}}glise, W~Del Pozzo, H~Dereli, R~De Rosa, L~Di Fiore, A~Di Lieto,
  A~Di Virgilio, M~Doets, V~Dolique, M~Drago, M~Ducrot, G~Endr{\H{o}}czi,
  V~Fafone, S~Farinon, I~Ferrante, F~Ferrini, F~Fidecaro, I~Fiori, R~Flaminio,
  J-D Fournier, S~Franco, S~Frasca, F~Frasconi, L~Gammaitoni, F~Garufi,
  M~Gaspard, A~Gatto, G~Gemme, B~Gendre, E~Genin, A~Gennai, S~Ghosh,
  L~Giacobone, A~Giazotto, R~Gouaty, M~Granata, G~Greco, P~Groot, G~M Guidi,
  J~Harms, A~Heidmann, H~Heitmann, P~Hello, G~Hemming, E~Hennes, D~Hofman,
  P~Jaranowski, R~J~G Jonker, M~Kasprzack, F~K{\'{e}}f{\'{e}}lian, I~Kowalska,
  M~Kraan, A~Kr{\'{o}}lak, A~Kutynia, C~Lazzaro, M~Leonardi, N~Leroy,
  N~Letendre, T~G~F Li, B~Lieunard, M~Lorenzini, V~Loriette, G~Losurdo,
  C~Magazz{\`{u}}, E~Majorana, I~Maksimovic, V~Malvezzi, N~Man, V~Mangano,
  M~Mantovani, F~Marchesoni, F~Marion, J~Marque, F~Martelli, L~Martellini,
  A~Masserot, D~Meacher, J~Meidam, F~Mezzani, C~Michel, L~Milano, Y~Minenkov,
  A~Moggi, M~Mohan, M~Montani, N~Morgado, B~Mours, F~Mul, M~F Nagy,
  I~Nardecchia, L~Naticchioni, G~Nelemans, I~Neri, M~Neri, F~Nocera, E~Pacaud,
  C~Palomba, F~Paoletti, A~Paoli, A~Pasqualetti, R~Passaquieti, D~Passuello,
  M~Perciballi, S~Petit, M~Pichot, F~Piergiovanni, G~Pillant, A~Piluso,
  L~Pinard, R~Poggiani, M~Prijatelj, G~A Prodi, M~Punturo, P~Puppo, D~S
  Rabeling, I~R{\'{a}}cz, P~Rapagnani, M~Razzano, V~Re, T~Regimbau, F~Ricci,
  F~Robinet, A~Rocchi, L~Rolland, R~Romano, D~Rosi{\'{n}}ska, P~Ruggi,
  E~Saracco, B~Sassolas, F~Schimmel, D~Sentenac, V~Sequino, S~Shah, K~Siellez,
  N~Straniero, B~Swinkels, M~Tacca, M~Tonelli, F~Travasso, M~Turconi,
  G~Vajente, N~van Bakel, M~van Beuzekom, J~F~J van~den Brand, C~Van~Den
  Broeck, M~V van~der Sluys, J~van Heijningen, M~Vas{\'{u}}th, G~Vedovato,
  J~Veitch, D~Verkindt, F~Vetrano, A~Vicer{\'{e}}, J-Y Vinet, G~Visser,
  H~Vocca, R~Ward, M~Was, L-W Wei, M~Yvert, A~Zadro {\.{z}}ny, and J-P Zendri.
\newblock Advanced virgo: a second-generation interferometric gravitational
  wave detector.
\newblock {\em Classical and Quantum Gravity}, 32(2):024001, dec 2014.

\bibitem{Mours_2006}
Beno{\^{\i}}t Mours, Edwige Tournefier, and Jean-Yves Vinet.
\newblock Thermal noise reduction in interferometric gravitational wave
  antennas: using high order {TEM} modes.
\newblock {\em Classical and Quantum Gravity}, 23(20):5777--5784, sep 2006.

\bibitem{Vinet_2007}
Jean-Yves Vinet.
\newblock Reducing thermal effects in mirrors of advanced gravitational wave
  interferometric detectors.
\newblock {\em Classical and Quantum Gravity}, 24(15):3897--3910, jul 2007.

\bibitem{PhysRevD.79.122002}
Simon Chelkowski, Stefan Hild, and Andreas Freise.
\newblock Prospects of higher-order laguerre-gauss modes in future
  gravitational wave detectors.
\newblock {\em Phys. Rev. D}, 79:122002, Jun 2009.

\bibitem{PhysRevD.82.012002}
Paul Fulda, Keiko Kokeyama, Simon Chelkowski, and Andreas Freise.
\newblock Experimental demonstration of higher-order laguerre-gauss mode
  interferometry.
\newblock {\em Phys. Rev. D}, 82:012002, Jul 2010.

\bibitem{PhysRevD.84.102002}
Charlotte Bond, Paul Fulda, Ludovico Carbone, Keiko Kokeyama, and Freise
  Andreas.
\newblock Higher order laguerre-gauss mode degeneracy in realistic, high
  finesse cavities.
\newblock {\em Phys. Rev. D}, 84:102002, Nov 2011.

\bibitem{PhysRevD.84.102001}
T.~Hong, J.~Miller, H.~Yamamoto, Y.~Chen, and R.~Adhikari.
\newblock Effects of mirror aberrations on laguerre-gaussian beams in
  interferometric gravitational-wave detectors.
\newblock {\em Phys. Rev. D}, 84:102001, Nov 2011.

\bibitem{Sorazu}
B~Sorazu, P~J Fulda, B~W Barr, A~S Bell, C~Bond, L~Carbone, A~Freise, S~Hild,
  S~H Huttner, J~Macarthur, and K~A Strain.
\newblock Experimental test of higher-order laguerre{\textendash}gauss modes in
  the 10 m glasgow prototype interferometer.
\newblock {\em Classical and Quantum Gravity}, 30(3):035004, jan 2013.

\bibitem{Freise_2004}
A~Freise, G~Heinzel, H~Lück, R~Schilling, B~Willke, and K~Danzmann.
\newblock Frequency-domain interferometer simulation with higher-order spatial
  modes.
\newblock {\em Classical and Quantum Gravity}, 21(5):S1067--S1074, feb 2004.

\bibitem{finesse}
Daniel~David Brown and Andreas Freise.
\newblock \textsc{Finesse}.
\newblock \url{http://www.gwoptics.org/finesse}, May 2014.
\newblock {The software and source code is available at
  \url{http://www.gwoptics.org/finesse}.}

\bibitem{Bond2017}
Charlotte Bond, Daniel Brown, Andreas Freise, and Kenneth~A Strain.
\newblock Interferometer techniques for gravitational-wave detection.
\newblock {\em Living Reviews in Relativity}, 19, Feb 2017.

\bibitem{PhysRevD.82.042003}
Jean-Yves Vinet.
\newblock Thermal noise in advanced gravitational wave interferometric
  antennas: A comparison between arbitrary order hermite and laguerre gaussian
  modes.
\newblock {\em Phys. Rev. D}, 82:042003, Aug 2010.

\bibitem{voyager}
R~X Adhikari, K~Arai, A~F Brooks, C~Wipf, O~Aguiar, P~Altin, B~Barr,
  L~Barsotti, R~Bassiri, A~Bell, G~Billingsley, R~Birney, D~Blair, E~Bonilla,
  J~Briggs, D~D Brown, R~Byer, H~Cao, M~Constancio, S~Cooper, T~Corbitt,
  D~Coyne, A~Cumming, E~Daw, R~deRosa, G~Eddolls, J~Eichholz, M~Evans, M~Fejer,
  E~C Ferreira, A~Freise, V~V Frolov, S~Gras, A~Green, H~Grote, E~Gustafson,
  E~D Hall, G~Hammond, J~Harms, G~Harry, K~Haughian, D~Heinert, M~Heintze,
  F~Hellman, J~Hennig, M~Hennig, S~Hild, J~Hough, W~Johnson, B~Kamai, D~Kapasi,
  K~Komori, D~Koptsov, M~Korobko, W~Z Korth, K~Kuns, B~Lantz, S~Leavey,
  F~Magana-Sandoval, G~Mansell, A~Markosyan, A~Markowitz, I~Martin, R~Martin,
  D~Martynov, D~E McClelland, G~McGhee, T~McRae, J~Mills, V~Mitrofanov,
  M~Molina-Ruiz, C~Mow-Lowry, J~Munch, P~Murray, S~Ng, M~A Okada, D~J Ottaway,
  L~Prokhorov, V~Quetschke, S~Reid, D~Reitze, J~Richardson, R~Robie,
  I~Romero-Shaw, R~Route, S~Rowan, R~Schnabel, M~Schneewind, F~Seifert,
  D~Shaddock, B~Shapiro, D~Shoemaker, A~S Silva, B~Slagmolen, J~Smith, N~Smith,
  J~Steinlechner, K~Strain, D~Taira, S~Tait, D~Tanner, Z~Tornasi, C~Torrie,
  M~Van Veggel, J~Vanheijningen, P~Veitch, A~Wade, G~Wallace, R~Ward, R~Weiss,
  P~Wessels, B~Willke, H~Yamamoto, M~J Yap, and C~Zhao.
\newblock A cryogenic silicon interferometer for gravitational-wave detection.
\newblock {\em Classical and Quantum Gravity}, 37(16):165003, jul 2020.

\bibitem{cosmicx}
Sheila Dwyer, Daniel Sigg, Stefan~W. Ballmer, Lisa Barsotti, Nergis Mavalvala,
  and Matthew Evans.
\newblock Gravitational wave detector with cosmological reach.
\newblock {\em Phys. Rev. D}, 91:082001, Apr 2015.

\bibitem{pykat}
DD~Brown and A~Freise.
\newblock Pykat, July 2017.
\newblock {\url{http://www.gwoptics.org/pykat}}.

\bibitem{brown2020pykat}
Daniel~D. Brown, Philip Jones, Samuel Rowlinson, Andreas Freise, Sean Leavey,
  Anna~C. Green, and Daniel Toyra.
\newblock Pykat: Python package for modelling precision optical
  interferometers.
\newblock 2020.
\newblock arXiv e-print 2004.06270, Submitted to SoftwareX.

\bibitem{Ast}
Stefan Ast, Sibilla~Di Pace, Jacques Millo, Mikhael Pichot, Margherita Turconi,
  and Walid Chaibi.
\newblock Generation of very high-order high purity gaussian modes via spatial
  light modulation.
\newblock 2019.

\bibitem{Heinze}
Henning~Vahlbruch Joscha~Heinze and Benno Willke.
\newblock Frequency-doubling of continuous laser light in the laguerre-gaussian
  modes lg00 and lg33.
\newblock {\em Opt. Lett.}

\bibitem{Jones}
Aaron~W. {Jones} and Andreas {Freise}.
\newblock Increased sensitivity of higher-order laser beams to mode mismatches.
\newblock {\em arXiv e-prints}, page arXiv:2007.12564, July 2020.

\bibitem{LMAcoatings}
C.~Michel, N.~Morgado, L.~Pinard, B.~Sassolas, R.~Bonnand, J.~Degallaix,
  D.~Forest, R.~Flaminio, and G.~Billingsley.
\newblock {Realization of low-loss mirrors with sub-nanometer flatness for
  future gravitational wave detectors}.
\newblock In Laurent Mazuray, Rolf Wartmann, Andrew~P. Wood, Marta~C. de~la
  Fuente, Jean-Luc~M. Tissot, Jeffrey~M. Raynor, Daniel~G. Smith, Frank
  Wyrowski, Andreas Erdmann, Tina~E. Kidger, Stuart David, and Pablo Benítez,
  editors, {\em Optical Systems Design 2012}, volume 8550, pages 516 -- 522.
  International Society for Optics and Photonics, SPIE, 2012.

\end{thebibliography}

\end{document}